\documentclass[11pt]{article}

\usepackage{cite}
\usepackage{wrapfig}
\usepackage{paralist}
\usepackage[margin=1in]{geometry}

\usepackage{subfigure}

\usepackage{url}

\usepackage[bf,small]{caption}
\usepackage{amssymb,amsmath,amsfonts,amsthm}

\usepackage{graphicx}
\usepackage{multirow,multicol}
\usepackage{xspace}
\usepackage{paralist}
\usepackage{color}
\usepackage{algpseudocode}
\usepackage{algorithm}

\usepackage{boxedminipage}
\usepackage{bibspacing}
\renewcommand{\paragraph}[1]{\vspace{5pt}\noindent\textbf{#1.}}
\newcommand{\ignore}[1]{}

\newcommand{\op}{\mathsf{op}}
\newcommand{\Read}{\mathsf{read}}
\newcommand{\Write}{\mathsf{write}}
\newcommand{\access}{\mathsf{Access}}
\newcommand{\data}{\mathsf{data}}
\newcommand{\slot}{\mathsf{slot}}
\newcommand{\evict}{\mathsf{Evict}}
\newcommand{\ecnt}{\mathsf{cnt}}
\newcommand{\seqevict}{\mathsf{SequentialEvict}}
\newcommand{\randevict}{\mathsf{RandomEvict}}

\newcommand{\pop}{\mathsf{pop}}
\newcommand{\fetchdel}{\mathsf{read\_and\_del}}

\newcommand{\step}{\mathsf{TimeStep}}

\newcommand{\pRead}{\mathsf{ReadPartition}}
\newcommand{\pWrite}{\mathsf{WritePartition}}
\newcommand{\cpRead}{\mathsf{ConcurrentReadPartition}}
\newcommand{\cpWrite}{\mathsf{ConcurrentWritePartition}}

\newcommand{\ind}{\mathsf{position}}

\newcommand{\N}{\mathbb{N}}

\newcommand{\Z}{\mathbb{Z}}
\newcommand{\R}{\mathbb{R}}

\newcommand{\cnt}{\mathsf{nextDummy}}
\newcommand{\prp}{\mathsf{PRP}}
\newcommand{\id}{\mathsf{u}}
\newcommand{\E}{\mathbb{E}}

\newcommand{\elaine}[1]{}
\newcommand{\emil}[1]{}

\newcommand{\defeq}{:=}

\newcommand{\etal}{\emph{et al.}\xspace}

\newcommand{\ceil}[1]{\ensuremath{\left\lceil {#1}\right\rceil}}
\newcommand{\floor}[1]{\ensuremath{\left\lfloor {#1}\right\rfloor}}

\newtheorem{definition}{Definition}
\newtheorem{remark}{Remark}

\newtheorem{theorem}{Theorem}
\newtheorem{lemma}{Lemma}

\newtheorem{fact}{Fact}
\newtheorem{proposition}{Proposition}
\newtheorem{observation}{Observation}

\def\final{1}
\ifnum\final=0
\newcommand{\mynote}[3]{\marginpar{\parbox{0.7in}{\tiny {\color{#2} {\sc #1}: {\sf #3}}}}}
\else
\newcommand{\mynote}[3]{}
\fi
\newcommand{\elainenote}[1]{\mynote{elaine}{blue}{#1}}

\begin{document}
%

\title{{\huge \bf Towards Practical Oblivious RAM}}

\author{
%
%
Emil Stefanov\\
       UC Berkeley\\
       emil@cs.berkeley.edu
\and
Elaine Shi\\
       UC Berkeley/PARC\\
       elaines@cs.berkeley.edu
\and
Dawn Song\\
       UC Berkeley\\
       dawnsong@cs.berkeley.edu
}

\date{}

\maketitle

\begin{abstract}
We take an important step forward in making Oblivious RAM (O-RAM) practical. We propose an O-RAM construction achieving an amortized overhead of $20$ $\sim$ $35$X (for an O-RAM roughly $1$ terabyte in size), about $63$ times faster than the best existing scheme. 
On the theoretic front, we propose
a fundamentally novel technique for constructing Oblivious RAMs:
specifically, we partition a bigger O-RAM into smaller O-RAMs, 
and employ a background eviction technique 
to obliviously evict blocks from the client-side cache into 
a randomly assigned server-side partition.
This novel technique is the key to achieving the gains in practical  
performance.
\end{abstract}

\section{Introduction}

\emil{Should we mention that our practical scheme is secure under the malicious model?}

\emil{Insert PDF metadata such as title and authors using the hyperref package.}

As cloud computing gains momentum, an increasing amount
of data is outsourced to cloud storage, and data privacy
has become an important concern for many businesses and individuals
alike. Encryption alone may not suffice for ensuring  
data privacy, as data access patterns can leak a considerable
amount of information about the data as well. 
Pinkas \etal gave an example~\cite{PinkasORAM}: 
if a sequence of data access requests $q_1, q_2, q_3$
is always followed by a stock exchange operation, the server
can gain sensitive information even when 
the data is encrypted.

Oblivious RAM (or O-RAM)~\cite{GSORAM,OsORAM,GoldORAM}, 
first investigated 
by Goldreich and Ostrovsky, is a primitive intended for hiding
storage access patterns. The problem was initially studied in the
context of software protection, i.e., hiding a program's 
memory access patterns
to prevent reverse engineering.

With the trend of cloud computing, 
O-RAM also has important applications in privacy-preserving
storage outsourcing applications. 
In this paper, we consider the setting where a client
wishes to store $N$ blocks each of size $B$ bytes at an untrusted server.

The community's interest in O-RAM has recently rekindled,
partly due to its potential high impact in 
privacy-preserving storage outsourcing applications. 
One of the best schemes known to date is a novel construction
recently proposed by Goodrich and Mitzenmacher~\cite{MMORAM}.
Specifically, 
let $N$ denotes the total storage
capacity of the O-RAM in terms of the number of blocks.
The Goodrich-Mitzenmacher construction achieves
$O((\log N)^2)$ amortized cost when parametrized 
with $O(1)$ client-side storage; or it achieves $O(\log N)$
amortized cost when parametrized with $O(N^a)$ client-side
storage where $0 < a < 1$. In this context, an amortized cost of $f(N)$ 
means that each data request will generate $f(N)$
read or write operations on the server.

Despite elegant asymptotic guarantees, the practical performance
of existing O-RAM constructions are still unsatisfactory.
As shown in Table~\ref{tab:contr}, one of the most practical schemes known to date is the construction by Goodrich and Mitzenmacher~\cite{MMORAM} when parametrized with $N^a$ ($a < 1$) client-side storage.
This scheme has more than $1,400X$
overhead compared to non-oblivious storage under reasonable 
parametrization, which is prohibitive in practice.  
In summary, although it has been nearly two decades
since Oblivious RAM was first invented,  
so far, it has mostly remained a theoretical concept.

\begin{table*}[t]
{\footnotesize
\begin{tabular}{|c|c|c|c|c|c|}
\hline
\multirow{2}{*}{Scheme} & \multirow{2}{*}{Amortized Cost} & \multirow{2}{*}{Worst-case Cost} & \multirow{2}{*}{Client Storage} & \multirow{2}{*}{Server Storage} & Practical \\ 
& & & & & Performance \\ \hline
\hline
Goldreich-Ostrovsky~\cite{GSORAM} & $O((\log N)^3)$ & $\Omega(N)$ & O(1) & $O(N \log N)$ & {\scriptsize $ > 120,000$X} \\ \hline
Pinkas-Reinman~\cite{PinkasORAM} & $O((\log N)^2)$ & $O(N \log N)$ & $O(1)$ & $8N$ & {\scriptsize $60,000 \sim 80,000$X} \\ \hline
\multirow{2}{*}{Goodrich-Mitzenmacher~\cite{MMORAM}} & \multirow{2}{*}{$O(\log N)$} & \multirow{2}{*}{$O(N \log N)$} & $O(N^a)$ & \multirow{2}{*}{$8N$} & \multirow{2}{*}{{\scriptsize $> 1,400$X}} \\
& & & ($0<a<1$) & & \\ \hline
\multicolumn{6}{|c|}{\bf This paper:}\\
\hline
Practical, Non-Concurrent& \multirow{1}{*}{$O(\log N)$} & \multirow{1}{*}{$< 3\sqrt{N} + o(\sqrt{N})$} & \multirow{1}{*}{$c N$ (c very small)} & \multirow{1}{*}{$< 4N + o(N)$} & \multirow{1}{*}{20 $\sim$ 35X}\\ \hline
Practical, Concurrent& \multirow{1}{*}{$O(\log N)$} & 
\multirow{1}{*}{$O(\log N)$} & 
\multirow{1}{*}{$c N$ (c very small)} & \multirow{1}{*}{$< 4N + o(N)$} & \multirow{1}{*}{20 $\sim$ 35X}\\ \hline
Theoretic, Non-Concurrent& \multirow{1}{*}{{\scriptsize $O((\log N)^2)$}} & \multirow{1}{*}{{\scriptsize $O(\sqrt{N})$}} & \multirow{1}{*}{$O(\sqrt{N})$}  & \multirow{1}{*}{$O(N)$} & \multirow{1}{*}{---}\\ 
\hline
Theoretic, Concurrent& \multirow{1}{*}{{\scriptsize $O((\log N)^2)$}} & \multirow{1}{*}{{\scriptsize $O((\log N)^2)$}} & \multirow{1}{*}{$O(\sqrt{N})$}  & \multirow{1}{*}{$O(N)$} & \multirow{1}{*}{---}\\ 
\hline
\end{tabular}
}
\caption{{\bf Our contributions. }
{
The practical performance is the number of client-server operations per O-RAM operation for typical realistic parameters, e.g., when the
server stores terabytes of data, and the client has several hundred
megabytes to gigabytes of local storage, and  $N \ge 2^{20}$. \emil{I think that we add latency to this table like Babis suggested. It's a very positive aspect of our practical construction.}
For our theoretic constructions, the same asymptotic bounds also work
for the more general case where client-side storage is $N^a$ for some constant $0 < a < 1$.  
}}
\label{tab:contr}
\end{table*}

\begin{table*}
\centering
{\footnotesize
\begin{tabular}{|l|c|c|c|c|c|c|} \hline
   \multirow{2}{*}{O-RAM Capacity}& \multirow{2}{*}{\# Blocks} & \multirow{2}{*}{Block Size} & \multirow{2}{*}{Client Storage} & \multirow{2}{*}{Server Storage} & \multirow{2}{*}{$\frac{\text{Client Storage}}{\text{O-RAM Capacity}}$}  & Practical \\
   & & & & & & Performance \\ \hline
   64 GB & $2^{20}$ & 64 KB & 204 MB & 205 GB & 0.297\% & 22.5X \\ \hline
   256 GB & $2^{22}$ & 64 KB & 415 MB & 819 GB & 0.151\% & 24.1X \\ \hline
   1 TB & $2^{24}$ & 64 KB & 858 MB & 3.2 TB & 0.078\% & 25.9X \\ \hline
   16 TB & $2^{28}$ & 64 KB & 4.2 GB & 51 TB & 0.024\% & 29.5X \\ \hline
   256 TB & $2^{32}$ & 64 KB & 31 GB & 819 TB & 0.011\% & 32.7X \\ \hline
   1024 TB & $2^{34}$ & 64 KB & 101 GB & 3072 TB & 0.009\% & 34.4X \\ \hline
\end{tabular}
}
\caption{Suggested parametrizations of our practical construction. {\normalfont The practical performance is the number of client-server operations per O-RAM operation as measured by our simulation experiments.}}
\label{tab:parameterizations}
\end{table*}

\subsection{Results and Contributions}
Our main goal is to make Oblivious RAM practical
for cloud outsourcing applications.

\paragraph{Practical construction}
We propose an Oblivious RAM construction
geared towards optimal practical performance.
The practical construction achieves an amortized overhead of \textbf{20 $\sim$ 35X} (Tables \ref{tab:contr} and \ref{tab:parameterizations}),
about \textbf{63 times faster} than the best known construction.
In addition, this practical construction also 
achieves \textit{sub-linear $O(\log N)$ worst-case cost}, and 
\textit{constant round-trip latency} per operation. 
Although our practical construction requires asymptotically 
linear amount of client-side storage, the constant is 
so small ($0.01\%$ to $0.3\%$ of the O-RAM capacity)
that in realistic settings, 
the amount of client-side storage is comparable to $\sqrt{N}$.

\vspace{2pt}
\noindent\textbf{Theoretical construction.}
By applying recursion to the practical construction, 
we can reduce the client-side storage to a sublinear amount, and
obtain a novel construction of theoretic interest, 
achieving $O((\log N)^2)$ amortized and \textit{worst-case} cost,
and requires $O(\sqrt{N})$ client-side storage, 
and $O(N)$ server-side storage.
Note that in the $O((\log N)^2)$ asymptotic notation,
one of the $\log N$ factors stems from the 
the depth of the recursion; and in realistic settings 
(see Table~\ref{tab:parameterizations}), 
the depth of the recursion is typically $2$ or $3$.

Table~\ref{tab:contr} summarizes our contributions 
in the context of related work. Table~\ref{tab:parameterizations} provides suggested parametrizations for our practical construction.

\subsection{Main Technique: Partitioning}

We propose a novel \textit{partitioning} technique, which
is the key to achieving the claimed theoretical bounds 
as well as major practical savings.
The basic idea is to partition a
single O-RAM of size $N$ blocks into $P$
different O-RAMs of size roughly $\frac{N}{P}$ blocks each. 
This allows us to break down a bigger
O-RAM into multiple smaller O-RAMs.

The idea of partitioning is motivated 
by the fact that the major source of overhead
in existing O-RAM constructions arises from an expensive
\textit{remote oblivious sorting} protocol performed between the client and the server.
Because the oblivious sorting protocol can take up to $O(N)$ time, existing
O-RAM schemes require $\Omega(N)$ time in the worst-case or have unreasonable $O(\sqrt N)$ amortized cost.

We partition the Oblivious RAM into roughly
$P = \sqrt{N}$ partitions, each having $\sqrt{N}$ blocks 
approximately. This way, the client can use $\sqrt{N}$ blocks of
storage to sort/reshuffle the data blocks 
locally, and then simply transfer the reshuffled
data blocks to the server. 
This not only circumvents the need for the expensive
oblivious sorting protocol, but also allows us
to achieve $O(\sqrt{N})$ worst-case cost.
Furthermore, by allowing reshuffling to happen concurrently with
reads, we can further reduce the worst-case cost 
of the practical construction to $O(\log N)$.

While the idea of partitioning is attractive, it 
also brings along an important challenge
in terms of security. Partitioning creates an extra channel through which 
the data access pattern can potentially be inferred by observing the sequence of partitions accessed.
Therefore, we must take care to ensure that
\textit{the sequence of partitions accessed does not leak information about
the identities of blocks being accessed}.
Specifically, our construction ensures that the sequence
of partitions accessed appears pseudo-random 
to an untrusted server.

It is worth noting that Ostrovsky and Shoup~\cite{OS97} also came up with 
a technique to spread the reshuffling work of 
the hierarchical solution~\cite{GSORAM}
over time, thereby achieving poly-logarithmic worst-case cost. 
However, 
our technique of achieving 
poly-logarithmic worst-case cost is fundamentally from 
Ostrovsky and Shoup~\cite{OS97}.
Moreover, our partitioning and background eviction techniques 
are also key to 
the practical performance gain that we can achieve.

\subsection{Related Work}
\label{sec:related}
Oblivious RAM was first investigated 
by Goldreich and Ostrovsky~\cite{GSORAM,OsORAM,GoldORAM}.
Since their original work, several seminal improvements
have been proposed~\cite{usablepir,Williams08,PinkasORAM,MMORAM}.
These approaches mainly fall into two broad categories:
constructions that use $O(1)$ client-side 
storage, and constructions that use $O(N^a)$ client-side storage
where $0 < a < 1$.

Williams and Sion~\cite{usablepir} propose an O-RAM construction
that requires $O(\sqrt{N})$ client-side storage, 
and achieves an amortized cost of $O((\log N)^2)$.
Williams \etal propose another construction that uses $O(\sqrt{N})$
client-side storage, and achieves $O(\log N \log \log N)$ 
amortized cost~\cite{Williams08}; however,
researchers have expressed concerns over the assumptions used
in their original analysis~\cite{PinkasORAM,MMORAM}. 
A corrected analysis of this construction can be found 
in an appendix in
a recent work by Pinkas and Reinman~\cite{PinkasORAM}.

Pinkas and Reinman~\cite{PinkasORAM} discovered an O-RAM construction
that achieves $O((\log N)^2)$ overhead with $O(1)$ 
client-side storage. However, 
some researchers have 
observed a security flaw of the Pinkas-Reinman construction,
due to the fact that the lookups
can reveal, with considerable probability, whether
the client is searching for blocks that  
exist in the hash table~\cite{MMORAM}.
The authors of that paper will fix 
this issue in a future journal version.
While Table~\ref{tab:contr} shows the overhead of the 
Pinkas-Reinman scheme as is, 
the overhead of the scheme is likely to increase after
fixing this security flaw.

In an elegant work by Goodrich and Mitzenmacher~\cite{MMORAM},
they proposed a novel O-RAM construction which
achieves $O((\log N)^2)$ amortized cost with $O(1)$ client-side
storage; or $O(\log N)$ amortized cost with $O(N^a)$ client-side
storage where $0 < a < 1$. 
The Goodrich-Mitzenmacher construction 
achieves the best asymptotic performance
among all known constructions.
However, their practical performance is still prohibitive.   
For example, with $O(\sqrt{N})$ client-side storage, 
their amortized cost is $>1,400$X from a very conservative estimate.
In reality, their overhead could be higher.
\elaine{See Appendix~\ref{sec:mmoram} for more details on how we obtained
this performance estimate. ... write that appendix}

In an independent and concurrent work 
by Boneh, Mazieres, and Popa~\cite{raluca},
they propose a construction that can support 
up to $O(\sqrt{N})$ reads while shuffling 
(using $O(\sqrt{N})$ client-side storage).
The scheme achieves 
$O(\log N)$ online cost, and $O(\sqrt{N})$ amortized cost. 

Almost all prior constructions
have $\Omega(N)$ worst-case cost, except the seminal work
by Ostrovsky and Shoup~\cite{OS97}, in which they demonstrate
how to spread the reshuffling operations of the hierarchical
construction~\cite{GSORAM}  
across time to achieve poly-logarithmic 
worst-case cost.
While the aforementioned concurrent work by Boneh \etal~\cite{raluca}
alleviates this problem by separating the cost into
an online part for reading and writing data, and an offline part
for reshuffling, they do so at an increased  
amortized cost of $O(\sqrt{N})$ 
(when their scheme is configured with $O(\sqrt{N})$ client-side storage). 
In addition, if $\Omega(\sqrt{N})$
consecutive requests take place within a small time window (e.g.,
during peak usage times),
their scheme can still block on a 
reshuffling operation of $\Omega(N)$ cost.

\paragraph{Concurrent and subsequent work}
In concurrent/subsequent work, 
Goodrich \etal~\cite{Goodrichworstcase} 
invented an O-RAM scheme achieving $O((\log N)^2)$
worst-case cost with $O(1)$ memory; and 
and Kushilevitz \etal~\cite{rafiworstcase} invented
a scheme with $O(\frac{(\log N)^2}{\log\log N})$
worst-case cost.
Goodrich \etal also came up with a stateless Oblivious 
RAM~\cite{statelessoram} scheme, with $O(\log N)$ amortized cost
and $O(N^a)$ ($0 < a < 1$) client-side 
transient (as opposed to permanent) buffers.
Due to larger constants in their constructions,
our construction is two to three orders of magnitude more efficient 
in realistic settings.

\section{Problem Definition}

\begin{figure*}
\begin{minipage}{0.48\textwidth}
\centering
\includegraphics[width=0.85\columnwidth]{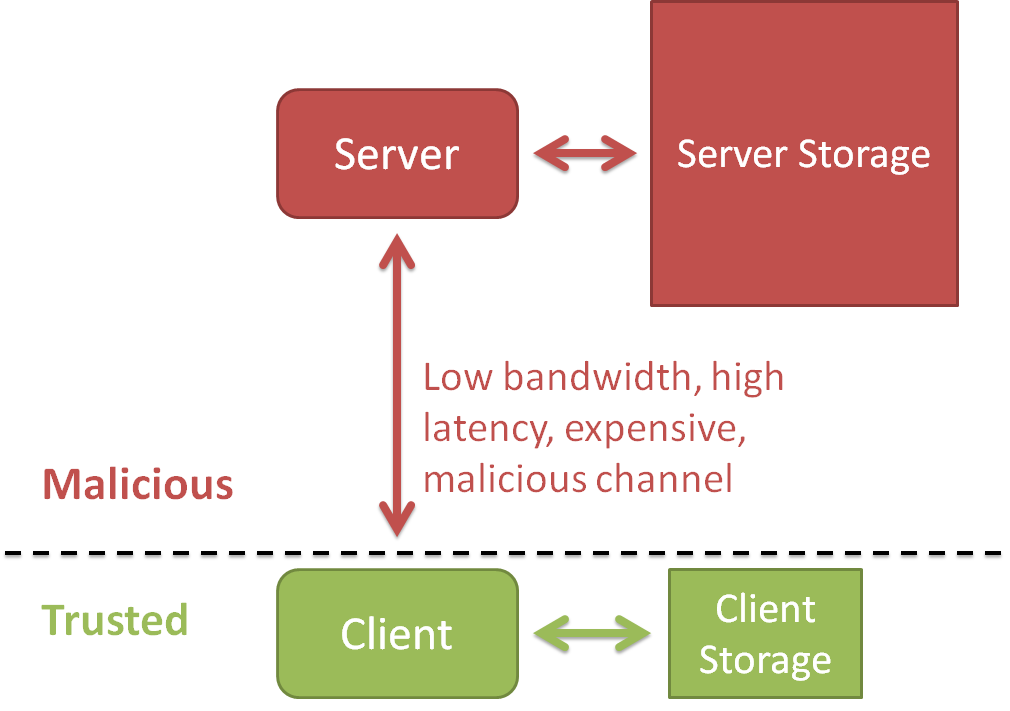}
\caption{\bf Oblivious RAM system architecture.}
\label{fig:ProblemDefinition}
\end{minipage}
\begin{minipage}{0.48\textwidth}
\centering
\includegraphics[width=\columnwidth]{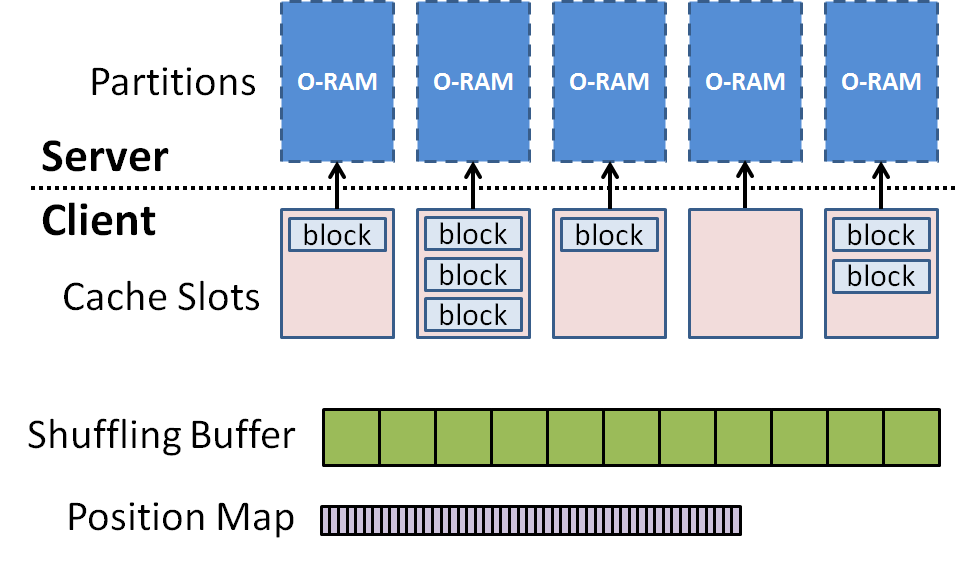}
\caption{{\bf The partitioning framework.}}
\label{fig:Partitioning}
\end{minipage}
\end{figure*}

As shown in Figure~\ref{fig:ProblemDefinition}, we consider a client
that wishes to store data at a remote untrusted server while preserving
its privacy. While traditional encryption 
schemes can provide confidentiality, they do not hide
the data access pattern which can reveal very sensitive
information to the untrusted server.
We assume that
the server is untrusted, and the client
is trusted, including the client's 
CPU and memory hierarchy (including RAM and disk).

The goal of O-RAM is to completely hide	
the data access pattern (which blocks were read/written) from the server.
In other words, each data read or write request 
will generate a completely random sequence of 
data accesses from the server's perspective.

\paragraph{Notations}
We assume that data is fetched and stored in atomic units, referred to as \textit{blocks}, of size $B$ bytes each. For example, a typical value for $B$ for cloud storage is 
$64$ KB to $256$ KB.
Throughout the paper, we use the notation $N$ to denote total number of data blocks that the O-RAM can support, also referred to as the capacity of the O-RAM. 

\paragraph{Practical considerations}
One of our goals is to design a \textit{practical} 
Oblivious RAM scheme in realistic settings.
We observe that 
\textit{bandwidth is much more costly 
than computation and storage} in real-world scenarios. 
For example, typical off-the-shelf PCs and laptops today have
gigabytes of RAM, and several hundred gigabytes of disk storage. 
When deploying O-RAM in a realistic setting, 
it is very likely that the bottleneck is network bandwidth and latency. 

As a result, our practical Oblivious RAM construction leverages 
available client-side storage as a working buffer, and this allows us to 
drastically optimize the bandwidth consumption between the server and the client.
As a typical scenario, we assume that the client wishes 
to store \textit{terabytes} of data
on the remote server, and the client has \textit{megabytes} to \textit{gigabytes}
of storage (in the form of RAM or disk). 
We wish to design a scheme in which the 
client can maximally leverage its local storage 
to reduce the overhead of O-RAM.

\paragraph{Security definitions}
We adopt the standard security definition for O-RAMs.
Intuitively, the security definition requires that the server
learns nothing about the access pattern. In other words,
no information should be leaked about:
1) which data is being accessed;
2) how old it is (when it was last accessed);
3) whether the same data is being accessed (linkability);
4) access pattern (sequential, random, etc);
or 5) whether the access is a read or a write.
Like previous work,  our O-RAM constructions do not consider information leakage through the timing channel, such as when or how frequently the client makes data requests. 

\begin{definition}[Security definition]
Let $\vec{y} := ((\op_1, \id_1, \mathsf{data}_1), (\op_2, \id_2, \mathsf{data}_2), ..., 
(\op_M, \id_M, \mathsf{data}_M))$
denote a data request sequence of length $M$, 
where each $\op_i$ denotes a 
$\Read(\id_i)$ or a $\Write(\id_i, \mathsf{data})$ operation.
Specifically, $\id_i$ denotes the identifier of the block being read or written, 
and $\mathsf{data}_i$ denotes the data being written.
Let $A(\vec{y})$ denote the (possibly randomized) sequence of 
accesses to the remote storage
given the sequence of data requests $\vec{y}$.
An O-RAM construction is said to be secure
if for any two data request sequences $\vec{y}$ and $\vec{z}$ of the same length, 
their access patterns 
$A(\vec{y})$ and $A(\vec{z})$ are computationally indistinguishable
by anyone but the client.  
\label{def:Obliviousness}
\end{definition}

\elaine{address the relaxed definition}
\emil{It would be convenient for the reader to know where we prove our scheme according to this definition.}

\emil{Talk about the malicious model.}

\section{The Partitioning Framework}
\label{sec:scheme}

In this section, we describe our main technique, \textit{partitioning}, as a framework. 
At a high level, the goal of partitioning is to subdivide the O-RAM into much smaller partitions, so that the operations performed on the partitions can be handled much more efficiently than if the O-RAM was not partitioned.

The main challenge of partitioning the O-RAM is to ensure that the sequence of partitions accessed during the lifetime of the O-RAM appears random to the untrusted server while keeping the client-side storage small. In this way, no information about data access pattern is revealed.

\subsection{Server Storage}

We divide the server's storage into $P$ fully functional \textit{partition O-RAM's}, each containing $N/P$ blocks on average. For now, we can think of each partition O-RAM as a black box, exporting a read and a write operation, while hiding the access patterns within that partition.

At any point of time, each block is randomly assigned to any of the $P$ partitions. Whenever a block is accessed, the block is logically removed from its current partition (although a stale copy of the block may remain), and logically assigned to a fresh random partition selected from all $P$ partitions. Thus, the client needs to keep track of which partition each block is associated with at any point of time, as specified in Section~\ref{sec:client}.

The maximum amount of blocks that an O-RAM can contain is referred to as the \textit{capacity} of the O-RAM. In our partitioning framework, blocks are randomly assigned to partitions, so the capacity of an O-RAM partition has to be slightly more than $N/P$ blocks to accommodate the variance of assignments. Due to the standard balls and bins analysis, 
for $P = \sqrt{N}$, 
each partition needs to have capacity $\sqrt{N} + o(\sqrt{N})$
to have a sufficiently small failure probability $\frac{1}{\text{poly}(N)}$.

\subsection{Client Storage}
\label{sec:client}

The client storage is divided into the following components.

\paragraph{Data cache with $P$ slots}
The data cache is a cache for temporarily storing
data blocks fetched from the server.
There are exactly $P$ cache slots, equal to the number 
of partitions on the server.
Logically, the $P$ cache slots can be thought of 
an extension to the server-side partitions. 
Each slot can store $0$, $1$, or multiple blocks.
In Appendix~\ref{sec:CacheBound}, we prove that
each cache slot will have a constant number of data blocks
in expectation, and that the total number of data blocks in all
cache slots
will be bounded by $O(P)$ with high probability
.
In both our theoretic and practical constructions,
we will let $P = \sqrt{N}$.
In this case, the client's data cache capacity is $O(\sqrt{N})$.

\paragraph{Position map}
As mentioned earlier, the client needs to keep track of which partition (or cache slot)
each block resides in. The position map serves exactly this purpose.
We use the notation $\ind[\id]$ for the partition number where block $\id$ currently resides.
In our practical construction described in Section~\ref{sec:partition}, 
the position map is extended to also contain the exact location (level number and index within the level) of block $\id$ within its current partition.

Intuitively, each block's position (i.e., partition number) requires about $c \log N$ bits to describe.
In our practical construction, $c \leq 1.1$, 
since the practical construction also stores the block's exact location inside a partition. 
Hence, the position map requires at most $c N \log N$ bits of storage, or $\frac{c}{8} N \log N$ bytes, which is $\frac{c N \log N}{8 \cdot B}$ blocks. Since in practice the block size $B > \frac{c}{8} \log N$, the size of the position map is a constant fraction 
of the original capacity of the O-RAM (with a very small constant).

\paragraph{Shuffling buffer}
The shuffling buffer is used for the shuffling operation
when two or more levels inside a partition O-RAM need to be merged
into the next level.
For this paper, we assume that the shuffling buffer has size
$O(\sqrt{N})$.

\paragraph{Miscellaneous}
Finally, we need some client-side storage to store miscellaneous
states and information, %
such as cryptographic keys for authentication, encryption, and pseudo-random permutations.
\subsection{Intuition}
In our construction, \textit{regardless of whether a block is found in the client's data cache, the client always performs a read and a write operation to the server upon every data request} -- with a dummy read operation in case of a cache hit. Otherwise, the server might be able to infer the age of the blocks being accessed. Therefore, the client data cache is required for security rather than for efficiency.

In some sense, the data cache acts like a holding buffer.
When the client fetches a block from the server, it cannot immediately
write the block back to a some partition, since this would 
result in linkability attacks the next the this block is read. Instead, the fetched block is associated with a fresh randomly chosen cache slot, but the block resides in the client data cache until a background eviction process writes it back to the server partition corresponding to its cache slot.

Another crucial observation is that the eviction process should not reveal which client cache slots are filled and which are not, as this can lead to linkability attacks as well. To achieve this, we use an eviction process that is independent of the load of each cache slot. Therefore we sometimes have to write back a dummy block to the server. For example, one possible eviction algorithm is to sequentially scan the cache slots at a fixed rate and evict a block from it or evict a dummy block if the cache slot is empty.

To aid the understanding of the partitioning framework, 
it helps to think of each client cache slot $i$ as an extension
of the $i$-th server partition.
At any point of time, 
a data block is associated with a random partition (or slot),
and the client has a position map to keep track of the location 
of each block. 
If a block is associated with partition (or slot) 
$i \in [P]$, it means that an up-to-date version of a block currently  
resides in partition $i$ (or cache slot $i$). However, it is possible
that other partitions (or even the same partition) 
may still carry a stale version of block $i$, which
will be removed during a future reshuffling operation. 

Every time a read or write operation is performed on a block, 
the block is re-assigned to a partition (or slot) 
\textit{selected independently at random from all 
$P$ partitions (or slots)}.
This ensures that two operations 
on the same block cannot be linked to each other.

\subsection{Setup}
When the construction is initialized, we first assign each block to an independently random partition. Since initially all blocks are zeroed, their values are implicit and we don't write them to the server. The position map stores an additional bit per block to indicate if it has never been accessed and is hence zeroed. In the practical construction, this bit can be implicitly calculated from other metadata that the client stores. Additionally, the data cache is initially empty.

\subsection{Partition O-RAM Semantics and Notations}
Before we present the main operations of 
our partitioning-based O-RAM, we first need to  
define the operations supported by each partition O-RAM.

Recall that each partition is a fully functional O-RAM by itself. To understand our partitioning framework, it helps to think of each partition as a blackbox O-RAM. For example, for each partition, we can plug in the Goodrich-Mitzenmacher O-RAM~\cite{MMORAM} (with either $O(1)$ or $O(\sqrt{N})$ client-side storage) or our own partition O-RAM construction described in Section~\ref{sec:partition}.

We make a few small assumptions about the 
partition O-RAM, and use slightly different semantics
to refer to the partition O-RAM operations 
than existing work. 
Existing O-RAM constructions~\cite{GoldORAM,PinkasORAM,MMORAM}
always perform both a read and a write operation 
upon any data access request.
For the purpose of the partitioning framework,
it helps to separate the reads from the writes. In particular,
we require that a $\pRead$ operation 
``logically remove'' the fetched block
from the corresponding partition. 
Many existing constructions~\cite{GoldORAM,PinkasORAM,MMORAM}
can be easily modified to support this operation,
simply by writing back a dummy block to
the first level of the hierarchy after reading.

Formally, we think of each partition O-RAM as a blackbox
O-RAM exporting two operations, 
$\pRead$ and $\pWrite$, as explained below.
\begin{compactitem}
\item
 $\pRead(p, \id)$ reads a block identified by its unique identifier $\id \in \{\bot, 1, 2, \ldots, N-1\}$
from partition $p$. In case $\id = \bot$, the read operation 
is called a dummy read. 
We assume that the $\pRead$ operation 
will \textit{logically} remove the fetched block
from the corresponding partition. 
\item
$\pWrite(p, \id, \mathsf{data})$ writes back a block identified by its unique identifier $\id \in \{\bot, 1, 2, \ldots, N-1\}$ to partition $p$. In case $\id = \bot$, the write operation is called a dummy write. The parameter $\mathsf{data}$ denotes the block's data.
\end{compactitem}

\begin{remark}[About the dummy block identifier $\bot$]
The dummy block identifier $\bot$ represents a meaningless data block. It is used as a substitute for a real block when the client does not want the server to know that there is no real block for some operation.
\end{remark}

\begin{remark}[Block identifier space]
Another weak assumption we make is that each partition O-RAM needs
to support non-contiguous block identifiers. In other words, the block
identifiers need not be a number within $[1, N]$, where $N$ is the O-RAM
capacity. Most existing schemes~\cite{MMORAM,PinkasORAM,GoldORAM} 
satisfy this property.
\end{remark}

\subsection{Reading a Block}
\label{sec:ops}

\begin{figure*}[t]
\centering
\subfigure{
\begin{boxedminipage}{0.9\textwidth}
\underline{$\access(\op, \id, \data^*)$}: 
\begin{algorithmic}[1]
\State $r \leftarrow \mathsf{UniformRandom}(1 \ldots P)$
\State $p \leftarrow \ind[\id]$, $\ind[\id] \leftarrow r$ \label{ln:AccessPositionMap}
\If{block $\id \textbf{ is in } \slot[p]$}
\State	$\data \leftarrow \slot[p].\fetchdel(\id)$
\State  $\pRead(p, \bot)$
\Else
\State  $\data \leftarrow \pRead(p, \id)$
\EndIf
\If{$\op = \Write$}
\State $\data \leftarrow \data^*$ 
\EndIf
\State $\slot[r] \leftarrow \slot[r] \cup \{ (\id, \data) \}$
\State Call $\evict(p)$ {\it \small \quad\quad /*Optional eviction piggy-backed on normal data access requests. Can improve performance by a constant factor.*/} \label{ln:piggybackevict}
\State Call $\seqevict(\nu)$ or $\randevict(\nu)$ \label{ln:backgroundevict}
\State \Return $\mathsf{\data}$
\end{algorithmic}
\end{boxedminipage}
}
\caption{{\bf Algorithm for data access}.
Read or write a data block identified by $\id$.
If $\op = \Read$, the input parameter $\data^* = \mathsf{None}$,
and the $\access$ operation returns the newly fetched block.
If $\op = \Write$, the $\access$ operation writes the specified
$\data^*$ to the block identified by $\id$, and returns
the old value of the block $\id$. 
}
\label{fig:access}
\end{figure*}

Let $\Read(\id)$ denote a read operation for a block identified by $\id$.
The client looks it up in the position map, and finds out which partition
block $\id$ is associated with. Suppose that block $\id$ is associated 
with partition $p$. The client then performs the following steps:

\vspace{2pt}
\noindent\textbf{Step 1}: Read a block from partition $p$.

\begin{compactitem}
\item[$\bullet$]
\noindent
If block $\id$ is found in cache slot $p$, the client performs a dummy read 
from partition $p$ of the server, i.e., call $\pRead(p, \bot)$ where $\bot$ denotes
a reading a dummy block. 

\item[$\bullet$]
Otherwise, the client reads block $\id$ from partition $p$ of the server by
calling $\pRead(p, \id)$.
\end{compactitem}

\vspace{2pt}
\noindent\textbf{Step 2}:
Place block $\id$ that was fetched in Step~1 into the client's cache, and update the position map.

\begin{compactitem}
\item
Pick a fresh random slot number $s$, and place block $\id$ into cache slot $s$. This means that block $\id$ is scheduled to be evicted to partition $s$ in the future, unless another $\Read(\id)$ preempts the eviction of this block.

\item
Update the position map, and 
associate block $\id$ with partition $s$.
In this way, the next $\Read(\id)$ 
will cause partition $s$ to be read and written.
\end{compactitem}

Afterwards, a background eviction takes place as described in Section~\ref{sec:BackgroundEviction}.

\subsection{Writing a Block}
Let $\Write(\id, \mathsf{data}^*)$ denote writing $\mathsf{data}^*$ to the block identified by $\id$.
This operation is implemented as a $\Read(\id)$ operation with the following exception: when block $u$ is placed in the cache during the $\Read(\id)$ operation, its data is set to $\mathsf{data}^*$.

\begin{observation}
Each $\Read$ or $\Write$ operation will  
cause an independent, random partition to be accessed. 
\label{obs:random}
\end{observation}
\begin{proof}(\textit{sketch.})
Consider each client cache slot as an extension of 
the corresponding server partition.
Every time a block $\id$ is read or written,
it is placed into a fresh random cache slot $s$, i.e.,
associated with partition $s$.
Note that every time $s$ 
is chosen at random, 
and independent of operations to the Oblivious RAM.
The next time block $\id$ is accessed, 
regardless of whether block $\id$ has been evicted 
from the cache slot before this access, 
the corresponding partition $s$ is read and written.
As the value of the random variable $s$ has not been revealed
to the server before this, from the server's perspective
$s$ is independently and uniformly at random. 
\end{proof}

\subsection{Background Eviction}
\label{sec:BackgroundEviction}
To prevent the client data cache from building up,
blocks need to be evicted to the server at some point.
There are two eviction processes: 
\begin{enumerate}
\item
\textit{Piggy-backed evictions} are those that take
place on regular O-RAM read or write operations (see
Line~\ref{ln:piggybackevict} of Figure~\ref{fig:access}).
Basically, if the data access request operates on a block
currently associated with partition $p$, we can piggy-back
a write-back to partition $p$ at that time.
The piggy-backed evictions are optional, but their
existence can improve performance by a constant factor.
\item
\textit{Background evictions} take place at a rate proportional to the data access rate (see Line~\ref{ln:backgroundevict} of Figure~\ref{fig:access}).
The background evictions are completely independent of the data access requests, and therefore can be equivalently thought of as taking place in a separate background thread. Our construction uses an eviction rate of $\nu > 0$, meaning that in expectation, $\nu$ number of background evictions are attempted with every data access request. Below are two potential algorithms for background eviction:
\begin{enumerate}
\item Sequentially scan the cache slots at a fixed rate $\nu$ (see the $\seqevict$ algorithm in Figure~\ref{fig:evict});
\item At a fixed rate $\nu$, randomly select a slot from all $P$ slots to evict from. (a modified version of the random eviction algorithm is presented as $\randevict$ in Figure~\ref{fig:evict}); \emil{We should really say why we present a modified version.}

\end{enumerate}

\end{enumerate}

Our eviction algorithm is designed to deal with two main challenges:

\begin{itemize}
\item \textbf{Bounding the cache size.} To avoid the client's data cache from building up indefinitely, the above two eviction processes combined evict blocks at least as fast as blocks are placed into the cache. The actual size of the client's data cache depends on the choice of the background eviction rate $\nu$. We choose $\nu > 0$ to be a constant factor of the actual data request rate. For our practical construction, in Section~\ref{sec:BandwidthExperiment} we empirically demonstrate the relationship of $\nu$ and the cache size. In Appendix~\ref{sec:CacheBound}, we prove that our background eviction algorithm results in a cache size of $O(P)$.

\item \textbf{Privacy.} It is important to ensure that the background eviction process does not reveal whether a cache slot is filled or the number of blocks in a slot. For this reason, if an empty slot is selected for eviction, a dummy block is evicted to hide the fact that the cache slot does not contain any real blocks.
\end{itemize}

\begin{observation}
By design, the background eviction process generates a
partition access sequence independent of the
data access pattern. 
\label{obs:evict}
\end{observation}

\begin{figure*}[t]
\subfigure{
\begin{boxedminipage}{0.4\textwidth}
\underline{$\evict(p)$}: 
\begin{algorithmic}[1]
\If{$\mathsf{len}(\slot[p])$ = 0}
\State $\pWrite(p, \bot, \mathsf{None})$
\Else
\State $(\id, \data) \leftarrow  \slot[\ecnt].\pop()$ 
\State $\pWrite(\ecnt, \id, \data)$ 
\EndIf
\end{algorithmic}
\end{boxedminipage}
}
\quad
\subfigure{
\begin{boxedminipage}{0.45\textwidth}
\underline{$\randevict(\nu)$}: 
\begin{algorithmic}[1]
\For{$i = 1$ to $\nu$} {\it \small \quad //Assume integer $\nu$}
\State $r \leftarrow \mathsf{UniformRandom(1 \ldots P)}$ 
\State $\evict(r)$
\EndFor
\end{algorithmic}
\end{boxedminipage}
}

\subfigure{
\begin{boxedminipage}{\textwidth}
\underline{$\seqevict(\nu)$}: 
\begin{algorithmic}[1]
\State $\mathsf{num} \leftarrow \mathcal{D}(\nu)$ {\footnotesize \it \quad \quad \quad
//Pick the number of blocks to evict according to distribution $\mathcal{D}$}
\For{$i = 1$ to $\mathsf{num}$}
\State $\ecnt \leftarrow \ecnt + 1$ 
{\small \it \quad \quad //$\ecnt$ is a global counter for the sequential scan}
\State $\evict(\ecnt)$
\EndFor
\end{algorithmic}
\end{boxedminipage}
}
\caption{
{\bf Background evicition algorithms with eviction rate $\nu$.}
Here we provide two candidate eviction algorithms
$\seqevict$ and $\randevict$.
$\seqevict$ determines the number of blocks to evict $\mathsf{num}$
based on a prescribed distribution $\mathcal{D}(\nu)$
and sequentially scans $\mathsf{num}$ slots to evict from.
$\randevict$ samples $\nu \in \N$ 
random slots (with replacement) to evict from.
{\bf In both $\seqevict$ and $\randevict$, if a slot selected for eviction
is empty, evict a dummy block for the sake of security.}
}
\label{fig:evict}
\end{figure*}

\begin{lemma}[Partition access sequence reveals nothing about the data request sequence.]
Let $\vec{y}$ 
denote a data request sequence.
Let 
$f(\vec{y})$ denote 
the sequence of partition numbers accessed
given data request sequence $\vec{y}$.
Then, for any two data request sequences of the same length 
$\vec{y}$ and $\vec{z}$, 
$f(\vec{y})$ and 
$f(\vec{z})$ are identically distributed.
In other words,
the sequence of partition numbers accessed during the life-time of the 
O-RAM does not 
leak any information
about the data access pattern.
\label{lem:partsec}
\end{lemma}
\begin{proof}
The sequence of partition numbers are generated in two ways
1) the regular $\Read$ or $\Write$ operations, and 
2) the background eviction processes.
Due to Observations~\ref{obs:random} and \ref{obs:evict},
both of the above processes generate a sequence of partition numbers  
completely independent of the data access pattern.
\end{proof}

\begin{theorem}
Suppose that each partition uses
a secure O-RAM construction, then the new O-RAM construction
obtained by applying the partitioning framework
over $P$ partition O-RAMs is also secure. \emil{This theorem is in the wrong subsection.}
\label{thm:PartitioningSecurity}
\end{theorem}
\begin{proof}
Straightforward conclusion from Lemma~\ref{lem:partsec}
and the security of the partition O-RAM.
\end{proof}

\subsection{Algorithm Pseudo-code}
Figures~\ref{fig:access}
and \ref{fig:evict}
describe in formal pseudo-code 
our Oblivious RAM operations based on the partitioning framework.
For ease of presentation, in Figure~\ref{fig:access}, we 
unify read and write operations into 
an $\access(\op, \id, \data^*)$ operation.

\section{Practical Construction}
\label{sec:partition}

In this section, 
we apply the partitioning techniques mentioned in the previous
section to obtain a practical construction 
with an amortized cost of $20 \sim 35$X overhead under typical settings,
about $63$ times faster than the best known construction. 
The client storage is typically $0.01\%$ to $0.3\%$ of the O-RAM capacity.
The worst-case cost of this construction is $O(\sqrt{N})$, and
we will later show how to allow concurrent shuffling and reads
to reduce the worst-case cost to $O(\log N)$ 
(Section~\ref{sec:concurrent}).

While the practical construction require $c N$  client-side
storage,  the constant $c$ is so small that 
our $c N$ is smaller than or comparable to $\sqrt{N}$
for typical storage sizes ranging from gigabytes to terabytes. 
For the sake of 
theoretic interest, in Appendix~\ref{sec:recursive}, we show
how to recursively apply our Oblivious RAM construction
to part of the client-side storage, and 
reduce the client-side storage to $O(\sqrt{N})$, while
incurring only a logarithmic factor in the amortized cost.

\subsection{Overview}

\begin{figure*}[t!]
\begin{minipage}{0.55\textwidth}
\begin{boxedminipage}{\columnwidth}
{\small
\underline{$\pRead(p, \id)$}:
\begin{algorithmic}[1]
\State $L \leftarrow$ number of levels
\For{$\ell = 0, 1, \ldots, L-1 \quad (\text{in parallel})$ }   
\If{level $\ell$ of partition $p$ is not filled}
\State continue   \quad \textit{// skip empty levels}
\EndIf
\If{block $\id$ is in partition $p$, level $\ell$}
\State $i = \ind[u].\text{index}$ 
\Else
\State $i = \ \cnt[p,\ell]$
\State $\cnt[p, \ell] \leftarrow \cnt[p, \ell] + 1$
\EndIf
\State $i' = \prp\left(K[p, \ell], i\right)$
\State Fetch from the server the block in partition $p$, level $\ell$, and
offset $i'$.
\State Decrypt the block with the key $K[p, \ell]$.
\EndFor
\end{algorithmic}
}

\end{boxedminipage}
\caption{
The $\pRead$ operation of our practical construction that reads the block with id $\id$ from partition $p$.
} 
\label{fig:ReadPartition}
\end{minipage}
\quad
\begin{minipage}{0.45\textwidth}
\centering
\includegraphics[width=0.56\columnwidth]{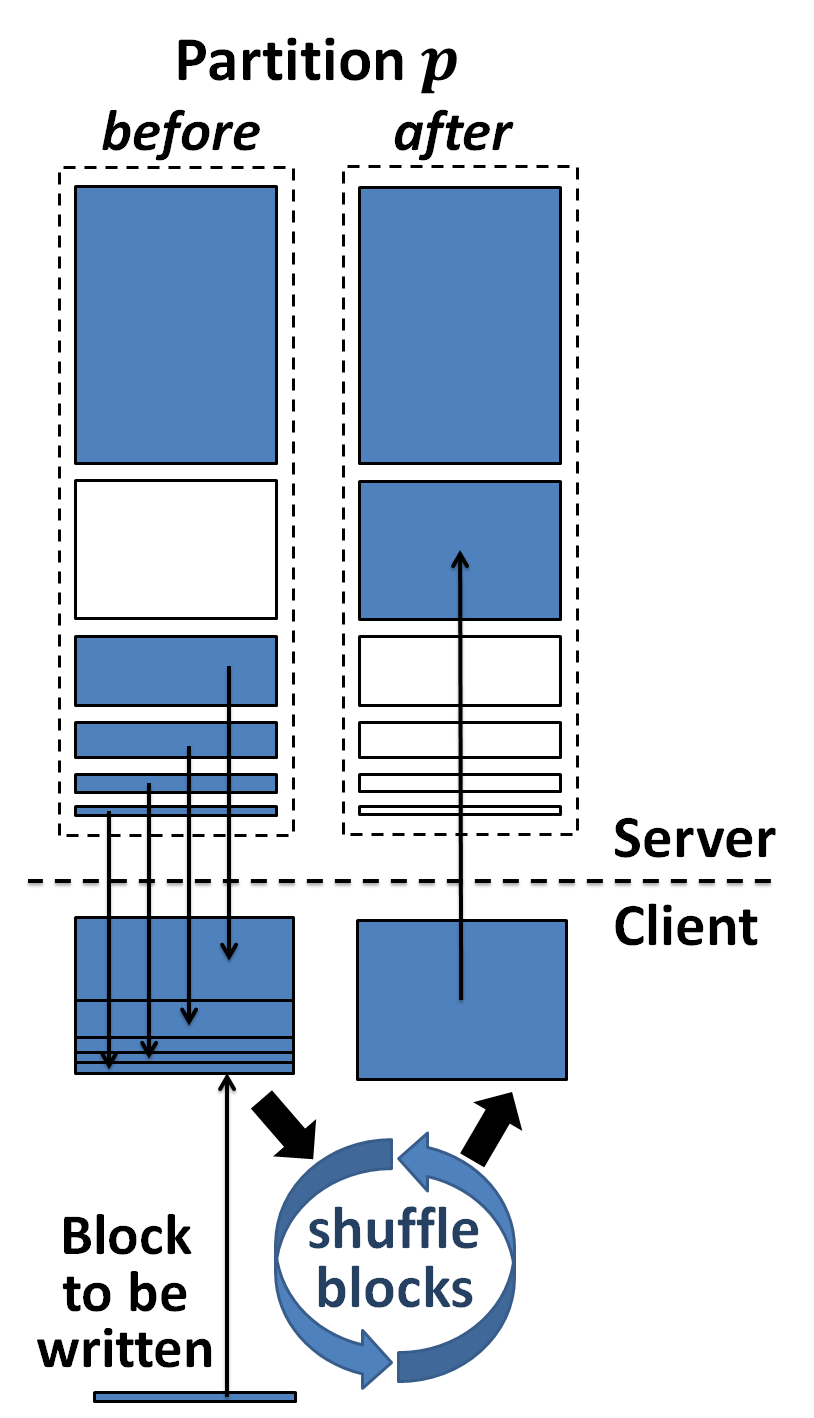}
\caption{$\pWrite$ leads to the shuffling of consecutively filled levels into the first empty level.
}
\end{minipage}
\end{figure*}

Our practical construction uses the partitioning framework (Section~\ref{sec:scheme}). For the partitions, we use our own highly optimized O-RAM construction resembling the Pinkas-Reinman
O-RAM at a very high level~\cite{PinkasORAM}. 

\paragraph{Choice of parameters} 
In this practical construction, we choose $P = \sqrt{N}$ partitions, each with $\sqrt{N}$ blocks on average. We use the $\seqevict$ algorithm as the background eviction algorithm. Every time 
$\seqevict$ is invoked with an eviction rate of $\nu$, it decides the number of blocks to evict $\mathsf{num}$based on a bounded geometric distribution with mean $\nu$, i.e., let $c$ be a constant representing the maximum number of evictions per data access operation, then $\Pr[\mathsf{num} = k] \propto p^k$ for $0 \leq k \leq c$, and $\Pr[\mathsf{num} = k] = 0$ for $k > c$.  Here is $0 < p < 1$ is a probability dependent on $\nu$ and $c$.

As mentioned earlier, the piggybacked evictions enable practical savings up to a constant factor, so we use piggybacked evictions in this construction.

\paragraph{Optimized partition O-RAM construction} 
While any existing O-RAM construction satisfying
the modified $\pRead$ and $\pWrite$
semantics can be used as a partition O-RAM,
we propose our own highly optimized partition O-RAM.
Our partition O-RAM construction resembles the Pinkas-Reinman O-RAM at a very high level~\cite{PinkasORAM}, but with several optimizations to gain practical savings. 
The practical savings come from at least three sources, in comparison with the Pinkas-Reinman construction:
\begin{compactitem}
\item
\textit{Local sorting.}
Due our partitioning framework, each partition is now of size $O(\sqrt{N})$ blocks. This allows us to use a client shuffling buffer of size $O(\sqrt{N})$ blocks to reshuffle the partition locally, 
thereby eliminating the need for extremely expensive oblivious sorting procedures during a reshuffling operation. 
This is our most significant source of saving in comparison
with all other existing schemes. 
\item
\textit{No Cuckoo hashing.}
Second, since we use a position map to save the locations of all blocks, we no longer need Cuckoo hashing, thereby saving a 2X factor for lookups.
\item
\textit{Compressed data transfer during reshuffling.}
Third, during the reshuffling operation, the client only reads blocks from each level that have not been previously read. Also, when the client writes back a set of shuffled blocks to the server (at least half of which are dummy blocks), it uses a compression algorithm to compress the shuffling buffer down to half its size. These two optimizations save about another 2X factor.
\item
\textit{Latency reduction.}
In the practical construction, the client saves a 
position map which records the locations of each 
block on the server. This allows the client to query the $O(\log N)$  
levels in each partition in a single round-trip, 
thereby reducing the latency to $O(1)$.
\end{compactitem}

\subsection{Partition Layout}
\label{sec:PartitionLayout}

As mentioned earlier, we choose $P = \sqrt{N}$ partitions for the practical construction. Each partition consists of $L = \log_2(\sqrt{N}) + 1 = \frac{1}{2} \log_2 N + 1$ levels, indexed by $0, 1, \ldots, \frac{1}{2} \log_2 N$ respectively. Except for the top level, each level $\ell$ has $2 \cdot 2^\ell$ blocks, among which at most half are real blocks, and the rest (at least half) are dummy blocks. \emil{Do we ever use capital $L$ in the rest of the paper? If not, then we should just remove it from here.}

The top level where $\ell = \frac{1}{2} \log_2 N$ has $2 \cdot 2^\ell + \epsilon = 2 \sqrt{N} + \epsilon$ blocks, where the surplus $\epsilon$ is due to the fact that some partition may have more blocks than others when the blocks are assigned in a random fashion to the partitions. 
Due to a standard balls and bins argument~\cite{ballsbins},
each partition's maximum size (including real and dummy blocks) 
should be $4\sqrt{N} + o(\sqrt{N})$
such that the failure probability $\frac{1}{\text{poly}(N)}$.
In Appendix~\ref{sec:maxpartload}, we empirically demonstrate that in practice, the maximum number of real blocks in each partition is not more than $1.15 \sqrt{N}$ for $N \ge 20$, hence the partition capacity is no more than $4 \cdot 1.15 \sqrt{N} = 4.6 \sqrt{N}$ blocks, and the total server storage is no more than $4.6N$ blocks. In Appendix~\ref{sec:ReducingServerStorage}, we propose an optimization to reduce the server storage to less than $3.2N$ blocks.

At any given time, a partition on the server might have some of its levels \textit{filled} with blocks and others \textit{unfilled}. The top partition is always filled. Also, a data block can be located in any partition, any filled level, and any offset within the level. In the practical construction, we extend the position map of the partition framework to also keep track of the level number and offset of each block.

From the perspective of the server, all blocks within a level are pseudo-randomly arranged. Because the blocks are encrypted, the server cannot even tell which blocks are real and which ones are dummy. We use keyed pseudo-random permutation ($\prp$) function for permuting blocks within a level in our construction. When the context is clear, we omit the range or the $\prp$ function in the pseudo-code.

\elainenote{can we delete the text in this setup paragraph
from ``Note that there are ...'' it doesnt seem to be too important}
\subsection{Setup}
The initial set of filled levels that contain the blocks depends on the partition number $p$. In order to better amortize the reshuffling costs of our scheme, the client randomly chooses which levels of each partition will be initially filled (with the restriction that the top level is always filled). Note that there are $2^{L-1}$ such possible \textit{fillings} of a partition where $L$ is the number of levels in the partition. The client notifies the server which levels are filled but does not write the actual blocks to the server because the blocks are initially zeroed and their values can be calculated implicitly by storing one bit for each level of each partition. This bit indicates if the entire level has never been reshuffled and is hence zeroed.

\subsection{Reading from a Partition}

The $\pRead$ operation reads the block with id $\id$ from partition $p$ as described in Figure~\ref{fig:ReadPartition}. If $\id = \bot$, then the $\pRead$ operation is a dummy read and a dummy block is read from each filled level. If $\id \ne \bot$, block $\id$ is read from the level that contains it, and a dummy block is read from from each of the other filled levels. Note that all of the fetches from the server are performed in parallel and hence this operation has single round trip latency unlike existing schemes \cite{GSORAM,OsORAM,GoldORAM} which take $\Omega(\log N)$ round trips.

\subsection{Writing to a Partition}
\label{sec:PracticalPartitionWrite}

\textit{Each write to a partition is essentially a reshuffling operation performed on consecutively filled levels in a partition.}
Therefore, we sometimes use the terms ``write'' and ``shuffling''
interchangeably.
 First, unread blocks from consecutively filled levels of the partition are read from the server into the client's shuffling buffer. Then, the client permutes the shuffling buffer according to a pseudo-random permutation (PRP) function. Finally, the client uploads its shuffling buffer into the first unfilled level and marks all of the levels below it as unfilled. The detailed pseudo-code for the $\pWrite$ operation is given in Figure~\ref{fig:WritePartition}.

There is an exception when all levels of a partition are filled. In that case, the reshuffling operation is performed on all levels, but at the end, the top level (which was already filled) is overwritten with the contents of the shuffling buffer and the remaining levels are marked as unfilled. Note that the shuffling buffer is never bigger than the top level because only \textit{unread real (not dummy)} blocks are placed into the shuffling buffer before it is padded with dummy blocks. Since the top level is big enough to contain all of the real items inside a partition, it can hold the entire shuffling buffer.

During a reshuffling operation, the client uses the pseudo-random permutation $\prp$ to determine the offset of all blocks (real and dummy) within a level on the server. Every time blocks are shuffled and written into the next level, the client generates a fresh random PRP key  $K[p, \ell]$ so that blocks end up at random offsets every time that level is constructed. The client remembers the keys for all levels of all partitions in its local cache.

\paragraph{Reading levels during shuffling} When the client reads a partition's levels into the shuffling buffer (Line~\ref{line:ReadUnreadBlocks} of Figure~\ref{fig:WritePartition}), it reads exactly $2^\ell$ previously unread blocks. Unread blocks are those that were written during a $\pWrite$ operation when the level was last constructed, but have not been read by a $\pRead$ operation since then. The client only needs to read the unread blocks because the read blocks were already logically removed from the partition when they were read. There is a further restriction that among those $2^\ell$ blocks must be all of the \textit{unread real (non-dummy)} blocks. Since a level contains up to $2^\ell$ real blocks, and there are always at least $2^\ell$ unread blocks in a level, this is always possible.

\elainenote{explain why you only need to read unread blocks}
The client can compute which blocks have been read/unread for each level. It does this by first fetching from the server a small amount of metadata for the level that contains the list of all blocks (read and unread) that were in the level when it was last filled. Then the client looks up each of those blocks in the position map to determine if the most recent version of that block is still in this level. Hence, the client can obtain the list of unread real blocks. The offsets of the of unread dummy blocks can be easily obtained by repeatedly applying the $\prp$ function to $\cnt$ and incrementing $\cnt$. Note that for security, the offsets of the $2^\ell$ unread blocks must be first computed and then the blocks must be read in order of their offset (or some other order independent of which blocks are real/dummy).

\begin{figure*}[t]
{\small
\begin{boxedminipage}{\linewidth}
\underline{$\pWrite(p, \id^*, \mathsf{data}^*)$}:
\begin{algorithmic}[1]
\State \textit{// Read consecutively filled levels into the client's shuffling buffer denoted \textsf{sbuffer}.}
\State $\ell_0 \leftarrow $ last consecutively filled level
\For{$\ell = 0$ to $\ell_0$}
\State Fetch the metadata (list of block ID's) for level $\ell$ in partition $p$. Decrypt with key $K[p, \ell]$.\label{line:FetchMetadata}
\State Fetch exactly $2^\ell$ previously unread blocks from level $\ell$ into \textsf{sbuffer} such that all unread real blocks are among them. Decrypt everything with the key $K[p, \ell]$. Ignore dummy blocks when they arrive. \label{line:ReadUnreadBlocks}
\State Mark level $\ell$ in partition $p$ as \textit{unfilled}. \label{line:MarkUnfilled}
\EndFor
\State $\ell = \min(\ell_0, L-1)$ \textit{// Don't spill above the top level.}
\vspace{0.6mm}
\State Add the $(\id^*, \mathsf{data}^*)$ to \textsf{sbuffer}.
\vspace{1mm} \label{line:AddDataToBuffer}
\State $k \leftarrow$ number of real blocks in \textsf{sbuffer}.
\For{$i = 1 $ to $k$ }
\State Let $(\id, \mathsf{data}) = \mathsf{sbuffer}[i]$ 
\State $\ind[u] \leftarrow \{p, \ell, i\}$ \textit{\quad // update position map}
\EndFor
\vspace{0.6mm}
\State $K[p, \ell] \leftarrow_R \mathcal{K}$ \textit{\quad\quad // generate fresh key for level $\ell$ in partition $p$}
\vspace{1mm}
\State Pad the shuffling buffer with dummy blocks up to length $2 \cdot 2^{\ell}$.
\textit{// The first $k$ blocks are real and the rest are dummy.}
\vspace{1mm}
\State Permute \textsf{sbuffer} with $\prp_\ell(K[p, \ell], \cdot)$. A block originally at index $i$ in the shuffling buffer is now located at offset $i'$ in the shuffling buffer, where $i' =\prp_\ell(K[p, \ell], i)$
\vspace{1mm}
\State Write the shuffling buffer into level $\ell$ in partition $p$ on the server, encrypted with key $K[p, \ell]$. \label{line:compression}
\State Write the metadata (list of block ID's) of level $\ell$ in partition $p$ to the server, encrypted with key $K[p, \ell]$. \label{line:WriteMetadata}
\State Mark level $\ell$ in partition $p$ as \textit{filled}.
\vspace{1mm}
\State $\cnt[p, \ell] \leftarrow k + 1$ \textit{\quad // initialize counter to first
dummy block.}
\end{algorithmic}

\vspace{1mm}

\underline{\bf Notations:}

\begin{tabular}{c|l}
$K[p, \ell]$  & Secret key for partition $p$, level $\ell$. (PRP or AES key depending on context) \\ 
$\cnt[p, \ell]$ &  Index of next unread dummy block for partition $p$, level $\ell$.\\ 
$\{p, \ell, i\} \leftarrow \ind[u]$  & {Position information for block $\id$ (partition $p$, level $\ell$, index $i$ within the level).}\\ 
\end{tabular}
\end{boxedminipage}
}
\caption{The $\pWrite$ operation of our practical construction that writes block $\id^*$ with $\mathsf{data}^*$ to partition $p$.}
\label{fig:WritePartition}
\end{figure*}

\subsection{Security}
Our practical construction has the following security guarantees: \textbf{obliviousness (privacy), confidentiality, and authentication (with freshness)} under the malicious model.

\begin{theorem}
The practical construction is oblivious according to Definition~\ref{def:Obliviousness}.
\end{theorem}
\begin{proof}
Theorem~\ref{thm:PracticalReadPartitionSecurity} and Theorem~\ref{thm:PracticalWritePartitionSecurity} in Appendix~\ref{sec:practicalsecproof} prove that the blocks accessed on the server via the $\pRead$ and $\pWrite$ operations are independent of the data request sequence. As can be seen in Figure~\ref{fig:access}, the order of the $\pRead$ and $\pWrite$ operations does not depend on the data request sequence, hence the construction is oblivious according to Definition~\ref{def:Obliviousness}.
\end{proof}

\begin{theorem}
The practical construction provides confidentiality and authentication (with freshness) of all data stored on a possibly malicious server. If the server corrupts or selectively modifies blocks in storage or in transit, the client can detect it.
\end{theorem}
\begin{proof}
This property is achieved by attaching a MAC to every block whenever a level is constructed (i.e., right after line~\ref{line:WriteMetadata} in Figure~\ref{fig:WritePartition}. The MAC is a function of the level's key $k$, the position $i$ of the block within the level, and the block ciphertext $x$.
$$m=\mathsf{MAC}_k (i,x)$$
Note that the level's key changes to a new random value every time the level is shuffled, so the same level key will never be reused in the future.
Whenever a block ciphertext $x$ in position $i$ of a level with key $k$ is read from the server storage, it is read along with its MAC $m$. Then, the block ciphertext is verified against the MAC (i.e., right after line~\ref{line:MarkUnfilled} in Figure~\ref{fig:WritePartition}) as follows:
$$m \stackrel{?}{=} \mathsf{MAC}_k(i,x)$$
The verification is done not only for real blocks, but also for dummy blocks stored within a level. For dummy blocks, the data x is always an encryption of all zeros.

We are able to do integrity checking and provide {\it freshness} guarantees by storing a single key for each filled level because levels are always constructed at once (during the atomic shuffle) and they are not modified until the level is deleted during its next reshuffling.
\end{proof}

It should be noted that although our partition O-RAM construction resembles the Pinkas-Reinman construction, it does not have the security flaw discovered by Goodrich and Mitzenmacher~\cite{MMORAM} because it does not use Cuckoo hash tables.

\section{Experimental Results}
\label{sec:experiment}

For our experiments, we implemented a simulator of our construction. Each read/write operation is simulated and the simulator keeps track of exactly where each block is located, the amount of client side storage used, and the total bytes transferred for all communication between the client and server. We also implemented a simulator for the best previously known O-RAM scheme for comparison.

\elainenote{say somewhere near here, in the experiments, we use bla bla optimizations
from appendix blabla}
For each parametrization of our Oblivious RAM construction, we simulated exactly $3N$ read/write operations. For example, for each O-RAM instances with $N=2^{28}$ blocks, we simulated about 800 million operations. We used a round-robin access pattern which maximizes the size of the client's data cache of our construction by maximizing the probability of a cache miss. Therefore our results always show the worst case cache size. Also, because our construction is oblivious, our amortized cost measurements are independent of the simulated access pattern. We used the level compression, server storage reduction, and piggy-backed eviction optimizations described in Appendix~\ref{sec:opt}.

\begin{figure*}[t!]
\begin{minipage}{0.48\textwidth}
\centering
\includegraphics[width=1\columnwidth]{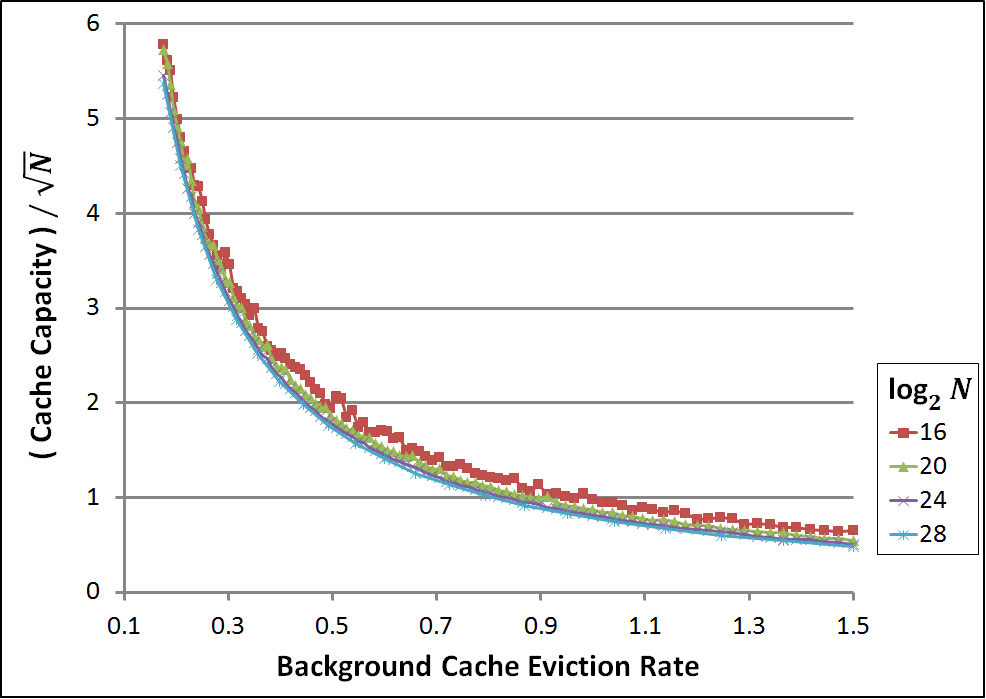}
\caption{{\bf Background Eviction Rate vs. Cache Capacity.} 
{\small
The $x$-axis is the eviction rate, defined as the ratio of background
evictions over real data requests. For example, an eviction
rate of $1$ suggests an equal rate of data requests and
background evictions. 
The $y$-axis is the 
quantity $(\textsf{cache capacity})/\sqrt{N}$, where
\textsf{cache capacity} is the maximum number of data blocks
in the cache over the course of time. 
}}
\label{fig:EvictionRate}
\end{minipage}
\begin{minipage}{0.48\textwidth}
\centering
\includegraphics[width=1\columnwidth]{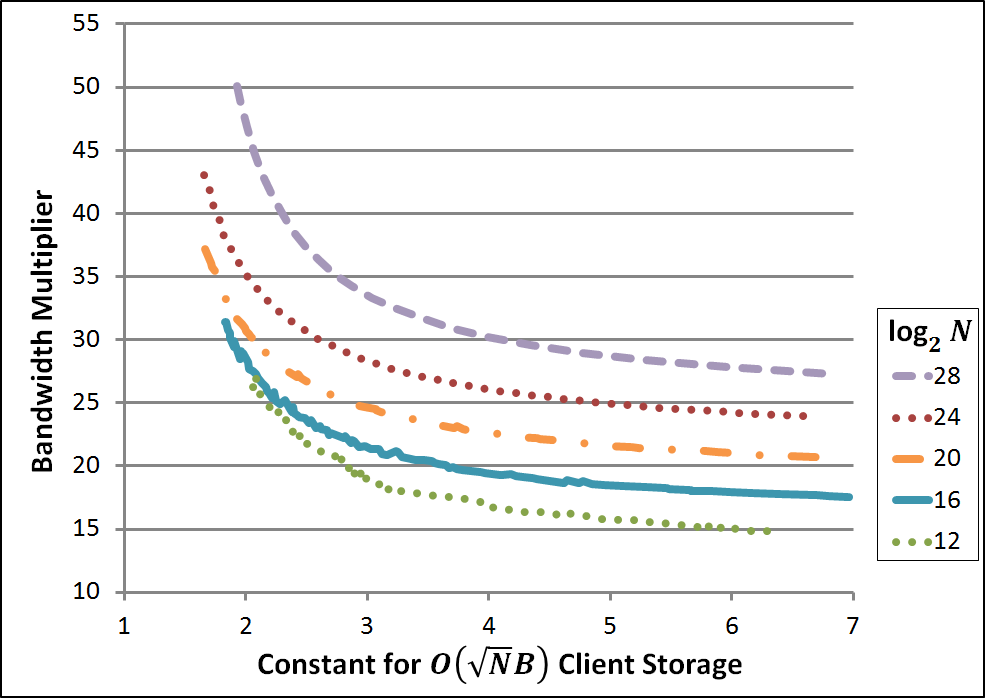}
\caption{{\bf Trade-off between Client Storage and Bandwidth.}  {\small The plot shows what bandwidth overhead a client can achieve by using exactly $k \sqrt{N}B$ bytes of client storage for different values of $k$ (horizontal axis). The client storage includes the cache, sorting buffer, and an uncompressed position map. A block size of 256 KB was assumed. \emil{Say ``practical performance''?}}}
\label{fig:ClientMemoryVsBandwidth}
\end{minipage}
\begin{minipage}{0.48\textwidth}
\centering
\includegraphics[width=1\columnwidth]{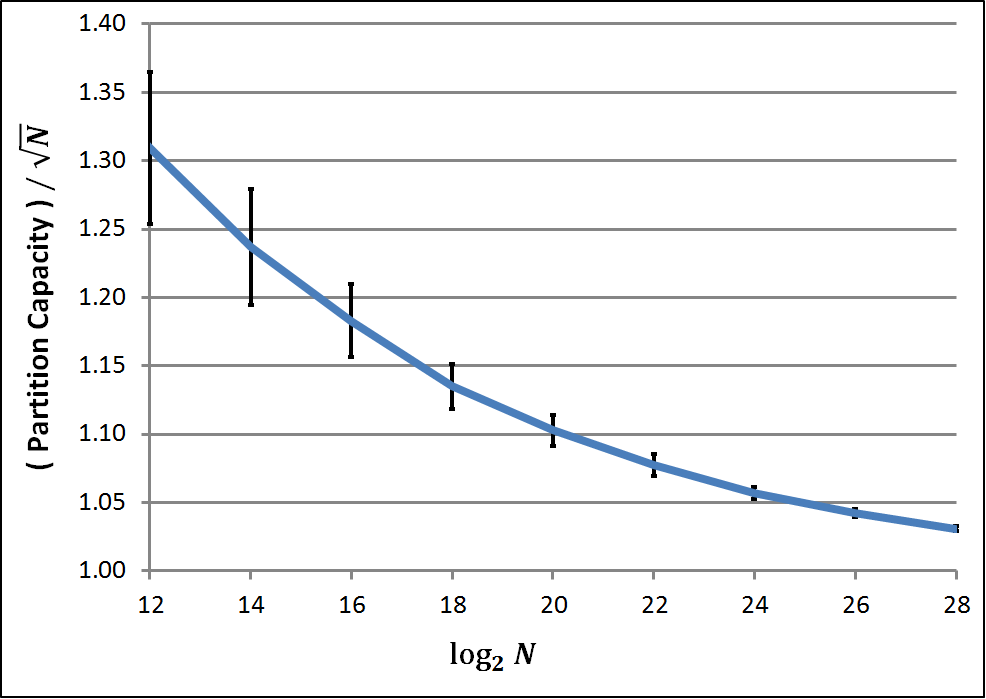}
\caption{{\bf Partition capacity.} {\small The $y$-axis is the quantity
$(\textsf{partition capacity})/\sqrt{N}$,
where \textsf{partition capacity} is the maximum number
of real data blocks that the partition must be able to hold,
i.e., the maximum number 
of real data blocks inside a partition over time.}%
}%
\label{fig:PartitionLoad}
\end{minipage}
\begin{minipage}{0.48\textwidth}
\centering
{\small
\begin{tabular}{|l|c|c|c|c|c|c|} \hline
\multirow{2}{*}{\# Blocks} & \multirow{2}{*}{Block Size}  & \multicolumn{2}{|c|}{Practical Performance} \\ \cline{3-4}
   & & Ours & Best Known\cite{MMORAM} \\ \hline
   $2^{16}$ & 16 MB & 18.4X & $> 1165$X \\ \hline
   $2^{18}$ & 4 MB & 19.9X & $> 1288$X \\ \hline
   $2^{20}$ & 1 MB & 21.5X & $> 1408$X \\ \hline
   $2^{22}$ & 256 KB & 23.2X & $> 1529$X \\ \hline
   $2^{24}$ & 64 KB & 25.0X & $>1651$X \\ \hline
\end{tabular}
}
\caption{{\bf Comparison between our construction and the best known previous O-RAM construction.} {\small A $1$ TB O-RAM is considered with both constructions using exactly $4 \sqrt{N}B$ client storage. The practical performance is the number of client-server operations per O-RAM operation. Our construction has a \textit{63 to 66 times better performance} than the best previously known scheme for the exact same parameters.}}
\label{tab:Comparison}
\end{minipage}
\begin{minipage}{0.48\textwidth}
\quad
\end{minipage}
\end{figure*}

\subsection{Client Storage and Bandwidth}
\label{sec:BandwidthExperiment}

In this experiment, we measure the performance overhead
(or \textit{bandwidth overhead}) of our O-RAM.
An O-RAM scheme with a bandwidth overhead of $w$ performs $w$ times the data transfer as an unsecured remote storage protocol. In the experiments we ignore the metadata needed to store and fetch blocks because in practice it is much smaller than the block size. For example, we may have 256 KB blocks, but the metadata will be only a few bytes.

In our scheme, the bandwidth overhead depends on the background eviction rate, and the background eviction rate determines the client's cache size. The client is free to choose its cache size by using the appropriate eviction rate. Figure~\ref{fig:EvictionRate} shows the correlation between the background eviction and cache size as measured in our simulation.

Once the client chooses its cache size it has determined the total amount of client storage. As previously mentioned, our scheme requires $O(\sqrt{N}B)$ bytes of client storage plus an extra $cNB$ bytes of client storage for the position map with a very small constant. For most practical values of $N$ and $B$, the position map is much smaller than the remaining  $O(\sqrt{N}B)$ bytes of client storage, so the client storage approximately scales like $O(\sqrt{N}B)$ bytes. We therefore express the total client storage as $k \sqrt{N}B$ bytes. Then, we ask the question: How does the client's choice of $k$ affect the bandwidth overhead of our entire O-RAM construction? Figure~\ref{fig:ClientMemoryVsBandwidth} shows this trade-off between the total client storage and the bandwidth overhead.

\subsection{Comparison with Previous Work}

To the best of our knowledge, the most practical existing O-RAM construction was developed by Goodrich~\etal~\cite{MMORAM}. It works by constructing a hierarchy of Cuckoo hash tables via Map-Reduce jobs and an efficient sorting algorithm which utilizes $N^a$ ($a < 1$) blocks of client-side storage. We implemented a simulator that estimates a lower bound on the performance of their construction. Then we compared it to the simulation of our construction.

To be fair, we parametrized both our and their construction to use the exact same amount of client storage: $4\sqrt{N}B$ bytes. The client storage includes all of our client data structures, including our position map (stored uncompressed). We parametrized both constructions for exactly 1 TB O-RAM capacity (meaning that each construction could store a total of 1 TB of blocks). We varied the number of blocks from $N=2^{16}$ to $N = 2^{24}$. Since the O-RAM size was fixed to 1 TB, the blocks size varied between $B = 2^{24}$ bytes and $B = 2^{16}$ bytes.
Table~\ref{tab:Comparison} shows the results. As it can be clearly seen, our construction uses \textit{63 to 66 times less bandwidth} than the best previously known scheme for the exact same parameters.

\subsection{Partition Capacity}
\label{sec:maxpartload}

Finally, we examine the effects of splitting up the Oblivious RAM into partitions. Recall that in our practical construction with $N$ blocks, we have split up the server storage into $\sqrt{N}$ partitions each containing about $\sqrt{N}$ blocks. Since the blocks are placed into partitions uniformly \textit{randomly} rather than uniformly, a partition might end up with slightly more or less than $\sqrt{N}$ blocks. For security reasons, we want to hide from the server how many blocks are in each partition at any given time, so a partition must be large enough to contain (with high probability) the maximum number of blocks that could end up in a single partition.

Figure~\ref{fig:PartitionLoad} shows how many times more blocks a partition contains than the expected number: $\sqrt{N}$. Note that as the size of the O-RAM grows, the maximum size of a partition approaches its expected size. In fact, one can formally show that the maximum number of real data blocks in each partition over time is
$\sqrt{N} + o(\sqrt{N})$~\cite{ballsbins}.
Hence, for large enough $N$, the partition capacity is less than $5\%$ larger than 
$\sqrt{N}$ blocks.

\section{Reducing the Worst-Case Cost With Concurrency}
\label{sec:concurrent}

The constructions described thus far have a worst-case cost $O(\sqrt N)$ because a $\pWrite$ operation sometimes causes a reshuffling of $O(\sqrt N)$ blocks.  We reduce the worst-case cost by spreading out expensive $\pWrite$ operations of $O(\sqrt N)$ cost over a long period of time, and at each time step performing $O(\log N)$ work. 

To achieve this, we allow reads and writes (i.e., reshuffling) to a partition to happen concurrently. This way, an operation does not have to wait for previous long-running operations to complete before executing. 
We introduce an \textit{amortizer} which keeps track of which partitions
need to be reshuffled, and schedules $O(\log N)$ work 
(or $O((\log N)^2)$ for the theoretic recursive construction) per time step.
There is a slight storage cost of allowing these operations to be done in parallel, but we will later \emil{reference} show that concurrency does not increase the asymptotic storage and amortized costs of our constructions. 

By performing operations concurrently, we decrease the worst-case cost of the practical construction from $O(\sqrt N)$ to $O(\log N)$ and we reduce the worst-case cost of the recursive construction from $O(\sqrt N)$ to $O((\log N)^2)$. 
Our concurrent constructions preserve the same amortized
cost as their non-concurrent counterparts;
however, in the concurrent constructions, the \textit{worst-case cost is 
the same as the amortized cost}. 
Furthermore, in the concurrent practical construction, the latency
is $O(1)$ just like the non-concurrent practical construction,
as each data request requires only a single round-trip to complete. 

\label{sec:concurrentdetail}
\subsection{Overview}
\begin{figure*}[t!]
\subfigure
{
\begin{minipage}{0.40\textwidth}
\includegraphics[width=1\textwidth]{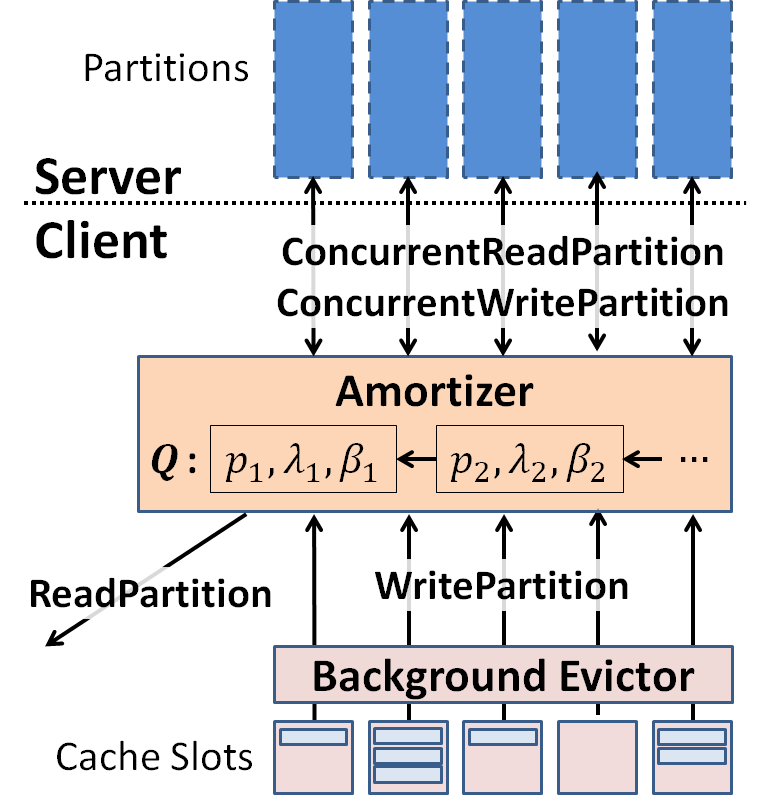}
\end{minipage}
}
\subfigure
{
\begin{boxedminipage}{0.55\textwidth}
\underline{$\step(\op)$}: 
\begin{algorithmic}[1]
\If{$op$ \textbf{is} $\pRead(p, \id)$}
\State Call $\cpRead(p, \id)$
\ElsIf{$op$ \textbf{is} $\pWrite(p, \id, \data)$}
\State $\lambda \leftarrow$ max $i$ for which $C_p = 0 \text{ mod } 2^i$.
\State $\beta \leftarrow \{(\id, \data)\}$
\If{$(p, \lambda', \beta') \in Q$ for some $\lambda'$ and $\beta'$}
\State $Q \leftarrow Q - \{(p, \lambda', \beta')\}$
\State $\lambda \leftarrow \max(\lambda, \lambda')$, $\beta \leftarrow \beta' \cup \beta$
\EndIf
\State $Q \leftarrow Q \cup \{(p, \lambda, \beta)\}$
\State $C_p \leftarrow C_p + 1$
\EndIf
\State Perform $O(\log N)$ work from job queue $Q$ in the form of $\cpWrite$ operations.
\end{algorithmic}
\end{boxedminipage}
}
\caption{{\bf The Amortizer component helps reduce the worst case costs of our constructions. }{It is inserted between the background eviction process and the server-side partitions as shown on the left. The component executes one operation per time step as defined on the right.}}
\label{fig:ShufflingAmortizing}
\end{figure*}

We reduce the worst case cost of our constructions by inserting an Amortizer component into our system which explicitly amortizes $\pRead$ and $\pWrite$ operations as described in Figure~\ref{fig:ShufflingAmortizing}. 
Specifically, the Amortizer schedules a $\pRead$ operation
as a $\cpRead$ operation, so the read can occur while shuffling.
A $\pRead$ always finishes in $O(\log N)$ time.
Upon a $\pWrite$ operation (which invokes the shuffling of
a partition), the Amortizer creates 
a new shuffling ``job'', and appends it to a queue $Q$ of jobs.
The Amortizer schedules $O(\log N)$ amount of work to be done
per \textit{time step} for jobs in the shuffling queue. 

If reads are taking place concurrently with shuffling, 
special care needs to be taken to avoid leakages through
the access pattern. This will be explained in the detailed scheme
description below.
\paragraph{Terminology}
To aid understanding, it helps to define the following terminology.
\begin{compactitem}
\item
\textbf{Job.}
A job $(p, \lambda, \beta)$ denotes a reshuffling of levels $0, \ldots, \lambda$ of partition $p$ and then writing the blocks in $\beta$ to partition $p$ on the server.

\item
\textbf{Job Queue $Q$.}
The \textit{job queue} $Q$ is a FIFO list of jobs. It is also possible to remove jobs that are not necessarily at the head of the queue for the purpose of merging them with other jobs, however jobs are always added at the tail.
\item
\textbf{Partition counter.}
Let $C_p \in \Z_{s}$ denote a counter for partition $p$, where $s$ is the maximum capacity of partition $p$. All operations on $C_p$ are modulus $s$.
\item
\textbf{Work.}
The \textit{work} of an operation is measured in terms of the number of blocks that it reads and writes to partitions on the server.
\end{compactitem}
\paragraph{Handling $\pRead$ operations}
The amortizer performs a $\pRead(p, \id)$ operation as a $\cpRead(p, \id)$ operation as defined in Sections~\ref{sec:concurrentread} for the practical and recursive constructions respectively. If block $\id$ is cached by a previous $\cpRead$ operation, then it is instead read from $\beta$ where $(p, \lambda, \beta) \in Q$ for some $\lambda$ and $\beta$.

\paragraph{Handling $\pWrite$ operations}
The amortizer component handles a $\pWrite$ operation by adding it to the job queue $Q$. The job is later dequeued in some time step and processed (possibly across multiple time steps). If the queue already has a job involving the same partition, the existing job is merged with the new job for the current $\pWrite$ operation. Specifically, if one job requires shuffling levels $0, \ldots, \lambda$ and the other job requires shuffling levels $0, \ldots, \lambda'$, we merge the two jobs into a job that requires shuffling levels $0, \ldots, \max(\lambda, \lambda')$. We also merge the blocks to be written by both jobs.

\paragraph{Processing jobs from the job queue}
For each time step, the reshuffling component perform $w \log N$ work for a predetermined constant $w$ such that $w \log N$ is greater than or equal to the amortized cost of the construction. Part of that work may be consumed by a $\cpRead$ operation executing at the beginning of the time step as described in Figure~\ref{fig:ShufflingAmortizing}. The remaining work is performed in the form of jobs obtained from $Q$.

\begin{definition}[Processing a job]
A job $(p, \lambda, \beta)$ is performed as a $\cpWrite(p, \lambda, \beta)$ operation that reshuffles levels $0, \ldots, \lambda$ of partition $p$ and writes the blocks in $\beta$ to partition $p$. The $\cpWrite$ operation is described in Sections~\ref{sec:concurrentwrite} for the practical and recursive constructions respectively. Additionally, every block read and written to the server is counted to calculate the amount of work performed as the job is running. A job may be paused after having completed part of its work.
\end{definition}

Jobs are always dequeued from the head of $Q$. At any point only a single job called the \textit{current job} is being processed unless the queue is empty (then there are no jobs to process). Each job starts after the previous job has completed, hence multiple jobs are never processed at the same time.

If the current job does not consume all of the remaining work of the time step, the the next job in $Q$ becomes the current job, and so on. the current job is \textit{paused} when the total amount of work performed in the time step is exactly $w \log N$. In the next time step, the current job is resumed from where it was paused.

We now explain how to perform $\cpRead$ and $\cpWrite$ operations in the practical construction to achieve an $O(\log N)$ worst-case cost with high probability. \emil{reference worst-case cost proof}

\subsubsection{Concurrent Reads}
\label{sec:concurrentread}
The client performs the $\cpRead(p, \id)$ operation by reading $0$ or $1$ blocks from each filled level $\ell$ of partition $p$ on the server as follows:

\begin{compactitem}
\item If level $\ell$ in partition $p$ contains block $\id$, then
	\begin{compactitem}
	\item Read block $\id$ from level $\ell$ in partition $p$ on the server like in the $\pRead$ operation. 
	\end{compactitem}
\item If level $\ell$ in partition $p$ does not contain block $\id$ and this level is not being reshuffled, then
	\begin{compactitem}
	\item Read the next dummy block from level $\ell$ in partition $p$ on the server like in the $\pRead$ operation.
	\end{compactitem}
\item If level $\ell$ in partition $p$ does not contain block $\id$ and this level is being reshuffled, then
	\begin{compactitem}
	\item 	Recall that when level $\ell$ is being reshuffled, $2^\ell$ previously unread blocks are chosen to be read. Let $S$ be the identifiers of that set of blocks for level $\ell$ in partition $p$.
	\item Let $S' \subseteq S$ be the ID’s of blocks in $S$ that were not read by a previous $\cpRead$ operation after the level started being reshuffled. The client keeps track of $S'$ for each level by first setting it to $S$ when a $\cpWrite$ operation begins and then removing $\id$ from $S'$ after every $\cpRead(p, \id)$ operation.
	\item If $S'$ is not empty, the client reads a random block in $S'$ from the server.
	\item If $S'$ is empty, then the client doesn't read anything from level $\ell$ in partition $p$. Revealing this fact to the server does not affect the security of the construction because the server already knows that the client has the entire level stored in its reshuffling buffer. 
	\end{compactitem}
\end{compactitem}

When $\id = \bot$, block $\id$ is treated as not being contained in any level. Due to concurrency, it is possible that a level of a partition needs to be read during reshuffling. In that case, blocks may be read directly from the client's shuffling buffer containing the level.

\subsubsection{Concurrent Writes}
\label{sec:concurrentwrite}
A $\cpWrite(p, \lambda, \beta)$ operation is performed like the non-concurrent $\pWrite$ operation described in Figure~\ref{fig:WritePartition}, except for three differences.

The first difference is that the client does not shuffle based on the last consecutively filled level. Instead it shuffles the levels $0, ..., \lambda$ which may include a few more levels than the $\pWrite$ operation would shuffle.

The second difference is that at Line~\ref{line:AddDataToBuffer} of Figure~\ref{fig:WritePartition}, the client adds all of the blocks in $\beta$ to the buffer.

The third difference is at Line~\ref{line:FetchMetadata} of Figure~\ref{fig:WritePartition}. In the non-concurrent construction, client fetches the list of $2^\ell$ blocks ID's in a level that is about to be reshuffled. It then uses this list to determine which blocks have already been read as described in Section~\ref{sec:PracticalPartitionWrite}). Because $2^\ell$ is $O(\sqrt N)$ fetching this metadata in the non-concurrent construction takes $O(\sqrt N)$ work in the worst case.

To ensure the worst case cost of the concurrent construction is $O(\log N)$, the metadata is stored as a bit array by the client. This bit array indicates which real blocks in that level have already been read. The client also knows which dummy blocks have been read because it already stores the $\cnt$ counter and it can apply the $\prp$ function for all dummy blocks between $k+1$ and $\cnt$ where $k$ is the number of real blocks in a level. Observe that the client only needs to store a single bit for each real block on the server. Hence this only increases the client storage by $2N + \epsilon\sqrt N$ bits, which is significantly smaller than the size of index structure that the client already stores.

\begin{theorem}[Practical concurrent construction]
With $1 - \frac{1}{\text{poly}(N)}$ probability, 
the concurrent practical construction described above
has $O(\log N)$ worst-case and amortized cost,
and requires $c N$ client-side storage with a very small $c$,
and $O(N)$ server-side storage.
\end{theorem}
\begin{proof}
The proof is in Appendix~\ref{sec:concurrentproof}.
\end{proof}

\section{Recursive Construction}
\label{sec:recursive}
\elaine{fix small n}

\emil{Choose if we want to call this the recursive or theoretical construction. It's confusing that it has two names.}

\begin{figure*}[t!]
\centering
\includegraphics[width=0.7\linewidth]{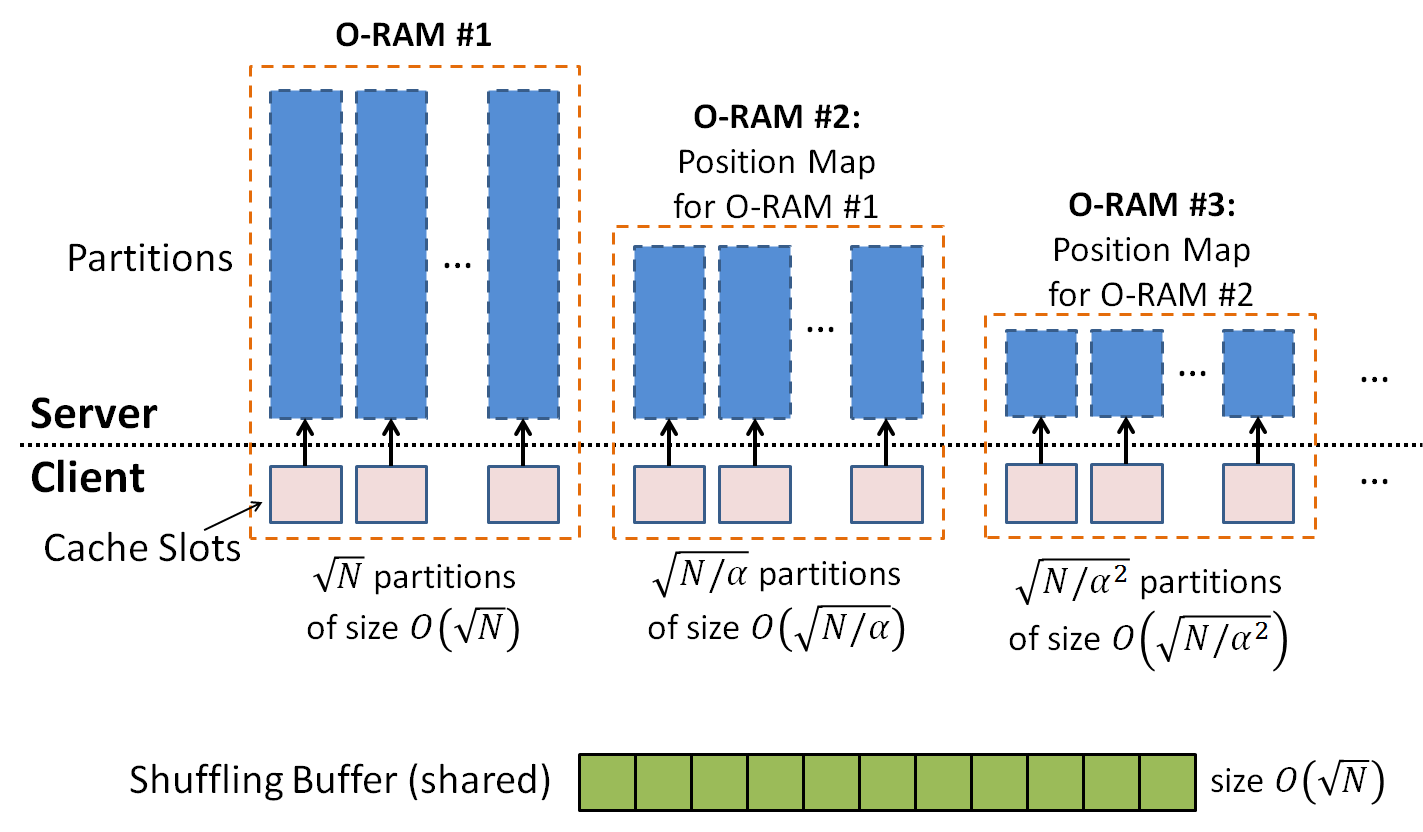}
\caption{\bf The recursive construction.}
\label{fig:RecursiveConstructionArchitecture}
\end{figure*}

The practical (non-current and concurrent) constructions 
described so far are geared towards optimal practical
performance. However, they are arguably not ideal in terms of asymptotic
performance, since they require a 
linear fraction of client-side storage 
for storing a position map of linear size.

For theoretic interest, we describe in this section
how to recursively apply our O-RAM constructions
to store the position map on the server,  
thereby obtaining O-RAM constructions with 
$O(\sqrt{N})$ client-side storage,
while incurring only a logarithmic factor in terms
of amortized and worst-case cost.

We first describe the recursive, non-concurrent construction
which achieves $O((\log N)^2)$ amortized cost, and
$O(\sqrt{N})$ worst-case cost.
We then describe how to apply the concurrency  
techniques to further reduce the worst-case cost
to $O((\log N)^2)$, such that the worst-case cost
and amortized cost will be the same.

\subsection{Recursive Non-Concurrent Construction}

\paragraph{Intuition}
Instead of storing the linearly sized position map locally, the client stores it in a separate O-RAM on the server. Furthermore, the O-RAM for the position map is guaranteed to be a constant factor smaller than the original O-RAM.  In other words, each level of recursion reduces the O-RAM capacity by a constant factor. After a logarithmic number of recursions, the size of the position map stored on the client is reduced to $O(1)$. The total size of all data caches is $O(\sqrt N)$, hence the construction uses $O(\sqrt N)$ client storage.

For the recursive construction, we employ the Goodrich-Mitzenmacher 
O-RAM scheme as the partition O-RAM. Specifically, we employ
their O-RAM scheme which for an O-RAM of capacity $N$,
achieves $O(\log N)$ amortized cost and 
$O(N)$ worst-case cost, while using $O(\sqrt{N})$ client-side storage,
and $O(N)$ server-side storage.

\begin{definition}[O-RAM$_{GM}$]
Let  O-RAM$_{GM}$
denote the Goodrich-Mitzenmacher
O-RAM scheme~\cite{MMORAM}: for an O-RAM of capacity $N$,
the O-RAM$_{GM}$ scheme achieves
$O(\log N)$ amortized cost, and 
$O(N)$ worst-case cost, while using $O(\sqrt{N})$ client-side storage,
and $O(N)$ server-side storage.
\end{definition}

\begin{definition}[O-RAM$_{base}$]
Let O-RAM$_{base}$ denote the O-RAM scheme derived through 
the partitioning framework with the 
following parameterizations: (1) we set $P = \sqrt{N}$ denote the number of partitions, where each
partition has approximately $\sqrt{N}$ blocks, and (2) we use the O-RAM$_{GM}$ construction as the partition O-RAM.
\end{definition}

Notice that in O-RAM$_{base}$, the client requires a 
data cache of size $O(\sqrt{N})$
and a position map of size less than $\frac{2 N \log N}{B}$ blocks.
If we assume that the data block size $B > 2 \log N$,   
then the client needs to store at most 
$ \frac{2 N \log N}{B} +  \sqrt{N} \log N = \frac{N}{\alpha} + o(N) $ 
blocks of data, 
where the compression rate 
$\alpha = \frac{B}{2 \log N} >  1$. 
To reduce the client-side storage, we can recursively apply the O-RAM construction to store the position map on the server side.

\begin{definition}[Recursive construction: O-RAM$^*$]
Let O-RAM$^*$ denote a recursive O-RAM scheme constructed as below. In O-RAM$_{base}$, the client needs to store a position map of size $cN (c < 1)$. Now, instead of storing the position map locally on the client, store it in a recursive O-RAM on the server side. The pseudocode of the O-RAM$^*$ scheme would be otherwise be the same as in Figure~\ref{fig:access} except that Line~\ref{ln:AccessPositionMap} is modified to the the following recursive O-RAM lookup and update operation:

The position $\ind[\id]$ is stored in block $\id / \alpha$ of the smaller O-RAM. The client looks up this block, updates the corresponding entry $\ind[\id]$ with the new value $r$, and writes the new block back. Note that the read and update can be achieved in {\bf a single O-RAM operation} to the smaller O-RAM. 
\end{definition}

\begin{theorem}[Recursive O-RAM construction]
Suppose that the block size $B > 2 \log N$ \emil{Don't we assume $B > 1.1 \log N$ now? If so, it should be changed here and everywhere else.}, and
that the number of data accesses $M < N^k$ for some
$k = O(\frac{\sqrt{N}}{\log N})$.
Our recursive O-RAM construction achieves 
$O((\log N)^2)$ amortized cost,
$O(\sqrt{N})$ worst-case cost, and requires
$O(N)$ server-side storage, and $O(\sqrt{N})$ client-side storage.
\label{thm:recursive}
\end{theorem}
\begin{proof}
The proof of the above theorem is presented in Appendix~\ref{sec:RecursiveConstructionCostsProofs}.
\end{proof}

\emil{Add a SECURITY proof for the recursive construction here.}

\subsection{Recursive Concurrent Construction}
Using similar concurrency techniques as 
in Section~\ref{sec:concurrent},
we can further reduce the worst-case cost of the recursive
construction to $O((\log N)^2)$.
Recall that the recursive construction differs from the practical construction in two ways: (1) it uses the O-RAM$_{GM}$ (Goodrich-Mitzenmacher \cite{MMORAM}) scheme as the partition O-RAM and (2) it recurses on its position map. We explain how to to perform concurrent operations in the O-RAM$_{GM}$ scheme to reduce the worst case cost of the base construction to $O(\log N)$ with high probability. Then when the recursion is applied, the recursive construction achieves a worst case cost of $O((\log N)^2)$ with high probability. \emil{Reference proofs.}

\subsubsection{Concurrent Reads}
As concurrency allows reshuffles to be queued for later, it is possible that a level $\ell$ is read more than $2^\ell$ times in between reshufflings. The O-RAM$_{GM}$ scheme imposes a restriction that at most $2^\ell$ blocks can be read from a level before it must be reshuffled by using a set of $2^\ell$ dummy blocks. We observe that it is possible to perform a \textit{dummy read} operation instead of using a dummy block and performing a \textit{normal read} on it.
This essentially eliminates the use of dummy blocks. Note that
the same idea was suggested in the work by Goodrich \etal~\cite{grouporam}. 

A dummy read operation $\cpRead(p, \bot)$ is performed by reading two random blocks within a level instead of applying a Cuckoo hash on a element from small domain. Observe that the Cuckoo hashes for real read operations output uniformly random block positions. Because the blocks read by dummy read operation are also chosen from a uniformly random distribution, dummy reads are indistinguishable from real reads.

This observation allows the client to securely perform a $\cpRead(p, \id)$ operation as follows. For each level (from smaller to larger) of partition $p$, as usual the client performs a Cuckoo hash of the block identifier $u$ to determine which two blocks to read within the level. Once the block is found in some level, the client performs dummy reads on subsequent levels. The client always first checks the local storage to see if the block is in the job queue. If so, then the client performs dummy reads of all levels.

In summary, instead of reading specific dummy blocks (which can be exhausted since there are only $2^\ell$ dummy blocks in level $\ell$), the client performs dummy reads by choosing two random positions in the level.

\subsubsection{Concurrent Writes}
In the concurrent recursive construction, the $\cpWrite(p, \lambda, \beta)$ operation is performed by first reshuffling levels $0, ..., \lambda$ along with the blocks in $\beta$ using the oblivious shuffling protocol of the O-RAM$_{GM}$ scheme. After the reshuffling has completed, the updated Cuckoo hash tables have been formed.

\begin{theorem}[Recursive concurrent O-RAM]
Assume that the block size $B > 2 \log N$.
With {\small $1 - \frac{1}{\text{poly}(N)}$} probability, 
the concurrent recursive construction described above
has $O((\log N)^2)$ worst-case and amortized cost,
and requires $O(\sqrt{N})$ client-side storage,
and $O(N)$ server-side storage.
\end{theorem}
\begin{proof}
The proof is in Appendix~\ref{sec:RecursiveConstructionCostsProofs}.
\end{proof}

\section{Optimizations and Extensions}
\label{sec:opt}
\label{sec:CompressingPositionMap}
\label{sec:ReducingServerStorage}
\subsection{Compressing the Position Map}
\label{sec:CompressingPositionMap}

The position map is \textit{highly} compressible under realistic workloads due to the natural sequentiality of data accesses. Overall we can compress the position map to about $0.255$ bytes per block. Hence its compressed size is $0.225 N$ bytes. Even for an extremely large, $1024$ TB O-RAM with $N = 2^{32}$ blocks, the position map will be under $1$ GB in size.
We now explain how to compress the position map.

\paragraph{Compressing partition numbers}
In \cite{OpreaCompression}, Oprea \etal showed that real-world file systems induce almost entirely sequential access patterns. They used this observation to compress a data structure that stores a counter of how many times each block has been accessed. Their experimental results on real-world traces show that the compressed data structure stores about $0.13$ bytes per block. Every time a block is accessed, their data structure stores a unique value (specifically, a counter) for that block. In our construction, instead of placing newly read blocks in a random cache slot, we can place them in a pseudo-random cache slot determined by the block id and counter. Specifically, when block $i$ is accessed for the $j$'th time (i.e., it's counter is $j$), it is placed in cache slot $\mathsf{PRF}(i.j)$. $\mathsf{PRF}(\cdot)$ is a pseudo-random function that outputs a cache slot (or partition) number.

\paragraph{Compressing level numbers} Recall that each partition contains $L$ levels such that level $\ell$ contains at most $2^\ell$ real blocks. We can represent the level number of each block by using only $1$ bit per block on average, regardless of the number of levels. This can be easily shown by computing the entropy of the level number as follows. If all levels are filled, each block has probability $2^{\ell-L}$ of being in level $\ell$. Then the entropy of the level number is
$$-\sum_{\ell=0}^{L-1}{2^{\ell-L} \log_2(2^{\ell-L}}) = \sum_{i=1}^{L}{i \cdot 2^{-i}} < 1$$
If not all levels in a partition are filled, then the entropy is even less, but for the sake of simplicity let's still use 1 bit to represent a level number within that partition. Note that since the top level is slightly larger (it contains $\epsilon$ extra blocks), the entropy might be slightly larger than $1$ bit, but only by a negligible amount.

\paragraph{Using the compressed position map} These two compression tricks allow us to compress our position map immensely. On average, we can use $0.13$ bytes for the partition number and $0.125$ bytes (1 bit) for the level number, for a total of $0.255$ bytes per block.

Once we have located a block's partition $p$ and level $\ell$, retrieving it is easy. When each level is constructed, each real block can be assigned a fresh alias $\mathsf{PRF}(K[p,\ell], \text{``real-alias''}, \id)$ where $\id$ is the ID of the block and $\mathsf{PRF}$ is a pseudo-random function. Each dummy block can be assigned the alias $\mathsf{PRF}(K[p,\ell], \text{``dummy-alias''}, i)$ where $i$ is the index of the dummy block in partition $p$, level $\ell$. Then during retrieval, the client fetches blocks from the server by their alias.

\subsection{Reducing Server Storage}
\label{sec:ReducingServerStorage}
Each partition's capacity is $\sqrt N + \epsilon$ blocks, where the surplus $\epsilon$ is due to the fact that some partitions may have more blocks than others when the blocks are assigned in a random fashion to the partitions. A partition has levels $\ell = 0, \ldots, \log_2{\sqrt{N}}$. Each level contains $2 \cdot 2^\ell$ blocks (real and dummy blocks), except for the top level that contains $2\epsilon$ additional blocks. Then, the maximum size of a partition on the server is $4\sqrt{N} + 2\epsilon$ blocks. Therefore, the maximum server storage is $4N + 2\epsilon\sqrt{N}$.

However, the maximum amount of server storage required is less than that, due to several reasons:

\begin{enumerate}
\item Levels of partitions are sometimes not filled. It is extremely unlikely that at some point in time, all levels of all partitions are simultaneously filled.
\item As soon as a block is read from a level, it can be deleted by the server because its value is no longer needed.
\end{enumerate}

In our simulation experiments, we calculated that the server never needs to store more than $3.2N$ blocks at any point in time. Hence, in practice, the server storage can be regarded as being less than $3.2N$ blocks.

\subsection{Compressing Levels During Uploading}
In Line~\ref{line:compression} of the $\pWrite$ algorithm in Figure~\ref{fig:WritePartition}, the client needs to write back up to $2\sqrt{N} + o(\sqrt{N})$ blocks to the server, at least half of which are dummy blocks. Since the values of the dummy blocks are irrelevant (since the server cannot differentiate between real and dummy blocks), it is possible to use a matrix compression algorithm
to save a 2X factor in terms of bandwidth.
Suppose that the client wishes to transfer  $2k$
blocks $b := (b_1, b_2, \ldots b_{2k})$. 
Let $S \subseteq \{1, 2, \ldots, 2k\}$ denote 
the offsets of the real blocks, 
let $b_S$ denote
the vector of real blocks.
We consider the case when exactly $k$ of the blocks are real,
i.e., $b_S$ is of length $k$ (if less than $k$ blocks
are real, simply select some dummy blocks to fill in).
To achieve the 2X compression, 
the server and client share a Vandermonde matrix $M_{2k \times k}$
during an initial setup phase.
Now to transfer the blocks 
$b$, the client solves the linear equation:
\[
M_{S}\cdot  x = b_S
\]
where $M_S$ denotes the matrix formed by rows of $M$ indexed by the set $S$, and
$b_S$ denote the vector $B$ indexed by the set $S$, i.e., the
list of real blocks.

The client can simply transfer $x$ (length $k$) to the server
in place of $b$ (length $2k$). 
The server decompresses it by computing $y \leftarrow M x $, and 
it is not hard to see that $y_S = b_S$.
The server is unable to distinguish which blocks are real and which
ones are dummy, since the Vandermonde matrix ensures that 
any subset of $k$ values of $y$ are a predetermined linear combination
of the remaining $k$ values of $y$.

\section*{Acknowledgments}

We would like to thank Hubert Chan, Yinian Qi, and Alina Oprea for insightful feedback,
helpful discussions, and proofreading.

This material is partially supported by the National Science Foundation Graduate Research Fellowship under Grant No. DGE-0946797, the National Science Foundation under Grants No. CCF-0424422, 0311808, 0832943, 0448452, 0842694, 0627511, 0842695, 0808617, 0831501 CT-L, by the Air Force Office of Scientific Research under MURI Award No. FA9550-09-1-0539, by the Air Force Research Laboratory under grant No. P010071555, by the Office of Naval Research under MURI Grant No. N000140911081, by the MURI program under AFOSR Grant No. FA9550-08-1-0352, and by a grant from the Amazon Web Services in Education program. Any opinions, findings, and conclusions or recommendations expressed in this material are those of the author(s) and do not necessarily reflect the views of the funding agencies. 

\emil{Do we have all of the references? It seems like tbe Binary O-RAM had a few more references.}
\emil{Cite Rafail Ostrovsky's paper with the worst-case costs.}

\bibliographystyle{abbrv}
\bibliography{refs}

\section*{\Large Appendices}
\appendix

\section{Bounding Partition Capacity and Cache Capacity}
\label{sec:proof}

\subsection{Bounding the Partition Capacity}
\label{sec:boundpartition}

\elaine{fix small N}

We now bound the partition capacity.
We will think of the $i$-th cache slot of the client
as an extension of the $i$-th partition on the server. 
We will bound the 
maximum number of data blocks in the partition plus corresponding cache slot. 
This is obviously
an upper-bound on the capacity of the corresponding partition.

We assume that the data access sequence is independent 
of the random coins used in the O-RAM scheme. 
Notice that this is also the case in practical applications.

Suppose we are given a data access sequence:
$$\vec{x} \defeq (x_1, x_2, \ldots, x_M).$$
If we think of the partition and its corresponding
cache slot as a unity, the operations of the O-RAM is equivalent
to the following random process:
Initially, each of the $N$ blocks is assigned to an 
independent random partion.
In each time step, pick an arbitrary element from its corresponding
partition, and place it in a fresh random partition.

Henceforth, when we perform probabilitic analysis on this random process,
we assume that the probability space is defined over the initial coin flips
that place each block into a partition, as well as the coin flips in
each time step $t$ that place $x_t$ into a random partition.

\begin{fact}
At any point of time, the distribution 
of blocks in the partitions (and their extended cache slots)
are the same as throwing $N$ balls into $\sqrt{N}$ bins.
\end{fact}

\begin{fact}
If we throw $s = N$ balls randomly into $t = \sqrt{N}$
bins,
\begin{align*}
\Pr[\text{a specific bin } > k \text{ balls} ] 
\leq  {s \choose k} \cdot \frac{1}{t^k} \leq \frac{s^k}{k!} \cdot \frac{1}{t^k} \leq
\left(\frac{\sqrt{N}}{k} \right)^k
\end{align*}
Therefore,
\begin{align*}
\Pr[\text{a specific bin } > \sqrt{N} + k \text{ balls} ] 
\leq \left(\frac{\sqrt{N}}{\sqrt{N} + k} \right)^{\sqrt{N} + k}
= \left(1 - \frac{1}{(\sqrt{N} + k)/k} \right)^{\sqrt{N} + k} \leq \frac{1}{e^k}
\end{align*}
\end{fact}

\begin{theorem}
Let $k > 0$.
At any time $t \in \N$, with probability $1 - o(1)$,
the number of blocks in the $i$-th partition $Z_i$ is bounded by 
$\sqrt{N} + k\ln N$,
i.e., 
\[
\Pr[Z_i > \sqrt{N} + k \ln N] 
\leq \frac{1}{N^k} 
\]
\label{thm:partsizesingle}
\end{theorem}

\begin{proof} (\textit{sketch}.)
Straightforward from the above calculation. \emil{Which calculation? Can we reference a specific proof?}
\end{proof}

\begin{theorem}[Partition capacity]
Let the total number of data requests $M \leq N^k$
for some positive $k$.
Given any sequence $\vec{x}$ of length $M$,
let $Z_{i, t}$ denote the load of the partition
$i$ at time step $t$. 
Then,
\[
\Pr\left[\exists i \in [N], t\in [M] : Z_{i,t} > \sqrt{N} + (k+c) \ln N \right]
\leq \frac{1}{N^c} 
\]
\label{thm:partsizesingle}
\label{thm:partitioncap}
\end{theorem}
\begin{proof}(\textit{sketch}.)
Due to Theorem~\ref{thm:partsizesingle}, and 
apply union bound over all partitions and all $M$ time steps.
\end{proof}

Given Theorem~\ref{thm:partsizesingle}, it suffices
to set each partition's capacity to be $\sqrt{N} + o(\sqrt{N})$. In other words,
each partition is an O-RAM that can store up to 
$\sqrt{N} + o(\sqrt{N})$ blocks.

\subsection{Bounding the Client Data Cache Size} 
\label{sec:CacheBound}

In the following analysis, we assume the $\randevict$ algorithm
with an eviction rate $\nu = 2$.

Recall that in our partitioning framework, eviction of blocks
from the data cache can happen in two ways: 1) piggy-backed evictions
that happen together with regular $\Read$ or $\Write$ operations;
and 2) background eviction. 
For the sake of upper-bounding the client data cache,
we pretend that 
there are no piggy-backed evictions, and only background evictions.
This obviously gives an upper bound on the data cache size.

\begin{figure}
\centering
\includegraphics[width=0.6\textwidth]{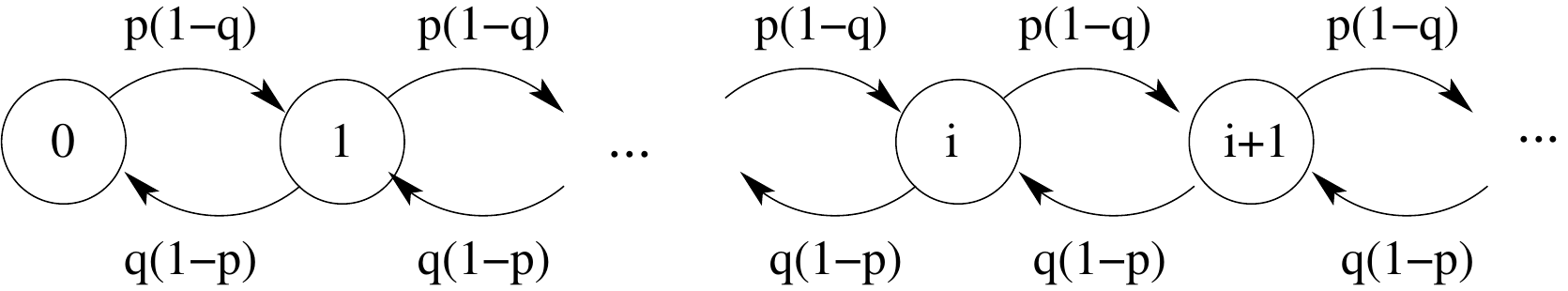}
\caption{Discrete-time Markov Chain for each cache slot.}  
\label{fig:mc}
\end{figure}

Now focus on a single slot. 
A slot can be viewed as a discrete-time Markov Chain
as shown in Figure~\ref{fig:mc}. Basically,
in every time step, with probability
$p = \frac{1}{\sqrt{N}}$, 
a block is added to the slot, and with probability
$q = \frac{2}{\sqrt{N}}$, a block is evicted (if one exists).

Let $\rho = \frac{p(1-q)}{q(1-p)} \leq \frac{1}{2}$.
Let $\pi_i$ denote the probability that a slot has 
$i$ blocks in the stationary distribution. Due to standard
Markov Chain analysis, we get: 
\[
\pi_i = \rho^i (1-\rho) 
\]  

\begin{fact}
At any given point of time, the expected number of
blocks in each slot is $\frac{\rho}{1-\rho}$,
and the expected number of blocks in all data cache slots
is $\frac{\rho P}{1-\rho}$, where $P$ is the number 
of partitions (or slots).
\label{fct:expcache}
\end{fact}

\begin{definition}[Negative association~\cite{negaassoc}]
A set of random variables $X_1, X_2, \ldots, X_k$ are negatively 
associated, if for every two disjoint index sets, 
$I, J \subseteq [k]$,
\[
\E[f(X_i, i \in I) g(X_j, j \in J)] \leq \E [ f(X_i, i \in I)] \E[g(X_j, j \in J)]
\]
for all functions $f: \R^{|I|} \rightarrow \R$
and $g:  \R^{|J|} \rightarrow \R$ that are both non-decresing
or both non-increasing.
\end{definition}

\begin{proposition}
Any any given point of time, let $Z_i$ denote the number
of blocks in slot $i \in [P]$. 
Then, the random variables $\{Z_i\}_{i \in [P]}$ are negatively
associated.
\end{proposition}
\begin{proof}
Let $B_{i,j}$ ($i \in [P], j\in[M]$) be indicator random variables defined as below:
\[
B_{i, j} = \begin{cases}
1 &  \text{if a block is placed in slot $i$ in the $j$-th time step} \\
0 &  \text{otherwise} \\
\end{cases}
\]

In each time step, $\nu$ slots will be randomly chosen for eviction.
Let $C_{i,j,k}$ ($i \in [P], j\in[M], k \in [\nu]$) be 
the indicator random variables defined as below:
\[
C_{i, j, k} = \begin{cases}
0 &  \text{if slot $i$ is selected for background eviction in time step $j$, $k$-th eviction} \\
1 &  \text{otherwise} \\
\end{cases}
\]

Notice that in the above, the indicator $C_{i,j,k}$ is $0$ 
(rather than $1$) if
slot $i$ is selected for background eviction in time step $j$.
This is an intentional choice which will later ensure that
the number of blocks in a slot is a non-decreasing
function of these indicator random variables.

Due to Proposition 11 of \cite{negaassoc}, the vector 
$\{B_{i,j}\}_{i \in [P], j \in [M]}$ is negatively associated.

Similarly, due to the Zero-One lemma of \cite{negaassoc},
the vector
$\{ 1 - C_{i,j,k}\}_{i \in [P], j \in [M], k \in [\nu]}$ is negatively associated,
and hence, the vector  
$\{C_{i,j,k}\}_{i \in [P], j \in [M], k \in [\nu]}$ is negatively associated,

Now, since all $B_{i, j}$ and 
$\{C_{i,j,k}\}$'s are mutually independent, 
the set of variables $\{B_{i,j}\}_{i \in [P], j \in [M]}
\cup \{C_{i,j,k}\}_{i \in [P], j \in [M], k \in [\nu]}$ 
is negatively associated due to 
Proposition 7 of \cite{negaassoc}.

For any $M \in \N$, after $M$ time steps,
the number of balls $Z_i$ in slot $i$ where $i \in [P]$
is a non-decreasing function of 
$Z_i \defeq Z_i(\{B_{i,j }\}_{j \in [M]}, 
\{C_{i,j,k }\}_{j \in [M], k \in [\nu]}
)$.
Due to Proposition 7 of \cite{negaassoc},
the number of blocks in each cache slot, i.e., the vector
$\{Z_i\}_{i \in [P]}$
is negatively associated.
\end{proof}

\begin{lemma}[Tail bound for sum of geometric random variables]
Let $X_1, X_2, \ldots, X_k$ be independent geometric random variables,
each having parameter $p$, and mean $\frac{1}{p}$. 
Let $X \defeq \sum_{i=1}^k X_i$, let $\mu = \frac{k}{p}$. Then,
for $0 <\epsilon < 1$, we have
\[
\Pr\left[X \geq (1+\epsilon) \mu \right] \leq 
\exp(-\frac{\epsilon^2 k}{4})
\]
\label{lem:chernoff}
\end{lemma}
\begin{proof}
Use the method of moment generating functions.
Suppose $e^t < \frac{1}{1-p}$.
\begin{align*}
\E[\exp(tX_i)]  = & p e^t + p (1-p) e^{2t} + p (1-p)^2 e^{3t} \ldots \\
=& \frac{p}{1-p} \sum_{i = 1}^\infty \left(e^t(1-p)\right)^i \\
= & \frac{p}{1-p}\cdot e^t (1-p) \cdot \frac{1}{1 - e^t(1-p)}\\
= & \frac{p e^t}{1-e^t(1-p)}
\end{align*}
\begin{align*}
\E[\exp(X)] = \prod_{i=1}^k \E[\exp(t X_i)] 
= \left(\frac{p e^t}{1-e^t(1-p)}\right)^k
= \left(\frac{p e^t}{1-e^t(1-p)}\right)^{\mu p}
\end{align*}

Therefore,
\begin{align*}
\Pr[X > (1+\epsilon) \mu] & = \Pr[\exp(tX) > \exp(t(1+\epsilon) \mu)] \\
& \leq 
\left(\frac{p e^t}{1-e^t(1-p)}\right)^{\mu p} \cdot \left( 
\frac{1}{e^{t(1+\epsilon)\mu}}
\right)\\
& = 
\left(\frac{p e^t}{1-e^t(1-p)}\right)^{\mu p} \cdot \left( 
\frac{1}{e^{t(1+\epsilon)/p}}\right)^{\mu p}\\
& = 
\left(\frac{p}{1-e^t(1-p)}\right)^{\mu p} \cdot \left( 
\frac{1}{e^{t(\frac{1 + \epsilon}{p} - 1)}}\right)^{\mu p}\\
& = 
\left(\frac{p}{1-e^t(1-p)} \cdot
\frac{1}{e^{t(\frac{1 + \epsilon}{p} - 1)}}\right)^{\mu p}\\
\end{align*}
The above inequality holds for all $t$ such that $e^t < \frac{1}{1-p}$.
We now pick the $t$ to minimize $\Pr[X > (1+\epsilon) \mu]$.
It is not hard to see that the 
above is minimized 
when $e^t = \frac{1 + \epsilon - p}{(1 + \epsilon)(1-p)}$.
Therefore, plugging 
$e^t = \frac{1 + \epsilon - p}{(1 + \epsilon)(1-p)}$ back in,
\begin{align*}
\Pr[X > (1+\epsilon) \mu] & \leq  
\left(\frac{p}{1-
\frac{1 + \epsilon - p}{1 + \epsilon}} \cdot
\frac{1}{   
\left(\frac{1 + \epsilon - p}{(1 + \epsilon)(1-p)}\right)^{(\frac{1 + \epsilon}{p} - 1)}}\right)^{\mu p}\\
& =  \left(
(1 + \epsilon) 
\left(\frac{(1 + \epsilon)(1-p)}{1 + \epsilon - p}\right)^{(\frac{1 + \epsilon}{p} - 1)}
\right)^{\mu p}\\
& =  \left(
(1 + \epsilon)^p
\left(1 - \frac{\epsilon p}{1 + \epsilon - p}\right)^{1 + \epsilon - p}
\right)^{\mu}\\
& \leq 
\left(
(1 + \epsilon)^p e^{-\epsilon p} 
\right)^\mu\\
& = \left( \frac{1+\epsilon}{e^\epsilon}\right)^k\\
\end{align*}

Specifically, for $0 < \epsilon < 1$, 
$\frac{1+\epsilon}{e^\epsilon} \leq \exp(-\frac{\epsilon^2 k}{4})$.  
Therefore, 
for $0 < \epsilon < 1$, we have:
\begin{align*}
\Pr[X > (1+\epsilon) \mu] \leq \exp\left(-\frac{\epsilon^2 k}{4}\right)
\end{align*}
\end{proof}

\begin{lemma}[Tail bound for sum of negatively-associated geometric random variables]
Let $X_1, X_2, \ldots, X_k$ be negatively 
associated geometric random variables,
each having parameter $p$, and mean $\frac{1}{p}$. 
Let $X \defeq \sum_{i=1}^k X_i$, let $\mu = \frac{k}{p}$. Then,
for $0 <\epsilon < 1$, we have
\[
\Pr\left[X \geq (1+\epsilon) \mu \right] \leq 
\exp(-\frac{\epsilon^2 k}{4})
\]
\end{lemma}
\begin{proof}
Similar to the proof of Lemma~\ref{lem:chernoff}.
Observe also that for negatively associated random variables
$X_1, X_2, \ldots, X_k$, for $t > 0$, we have
\[
\E[\exp(tX)] = \E[\prod_{i = 1}^k \exp(tX_i)]
\leq \prod_{i = 1}^k \E[\exp(tX_i)]
\]
The remainder of the proof follows directly from 
the proof of Lemma~\ref{lem:chernoff}.
\end{proof}

\begin{fact}
At any given point of time, let $Z_i$ denote the number
of blocks in slot $i \in [P]$, where $P$ is the number 
of partitions (or slots).
Then, $Y_i \defeq Z_i + 1$ is a geometrically distributed random variable
with mean $\frac{1}{1-\rho}$.
Moreover, as the random variables $\{Z_i\}_{i \in [P]}$
are negatively associated, so are 
$\{Y_i\}_{i \in [P]}$.
\end{fact}

\begin{lemma}
Let $Z$ denote the total cache size at any point of time
after the chain has reached stationary distribution. 
Let $P = \sqrt{N}$ denote the number of partitions (or slots).
From Fact~\ref{fct:expcache}, 
$\E[Z] = \frac{\rho P}{1-\rho} \leq \sqrt{N}$.
We have the following tail bound:
\[
\Pr[Z > \sqrt{N} + 4c N^{\frac{1}{4}} \sqrt{\ln N} ]  \leq \frac{1}{N^{c^2}}
\]  
\label{lem:cachesingletime}
\end{lemma}

\begin{proof}
Let $Y =\sum_{i = 1}^P Y_i = \sum_{i = 1}^P (Z_i+1)$  
denote the sum of negatively associated geometric random variables
$Y_i$, with mean $\frac{1}{1-\rho}$.
Therefore $\E[Y]= \frac{\sqrt{N}}{1-\rho} \leq 2\sqrt{N}$.
 
\begin{align*}
\Pr[Z > \E[Z] + 4c N^{\frac{1}{4}} \sqrt{\ln N}]
= \Pr[Y > \E[Y] + 4c N^{\frac{1}{4}} \sqrt{\ln N}]
\end{align*}
Let $$\epsilon \defeq 
\frac{4c N^{\frac{1}{4}} \sqrt{\ln N}}{\E[Y]} \geq
\frac{2c \sqrt{\ln N}}{N^{\frac{1}{4}}}$$
Notice that $0 < \epsilon < 1$ for sufficiently large
$N$. Therefore,
\begin{align*}
& \Pr[Z > \E[Z] + 4c N^{\frac{1}{4}} \sqrt{\ln N}]\\
= &\Pr[Y > \E[Y] + 4c N^{\frac{1}{4}} \sqrt{\ln N}]\\ 
= & \Pr[Y > (1 + \epsilon) \E[Y]]
\leq \exp(-\frac{\epsilon^2 \sqrt{N}}{4}) 
\leq \exp(-\frac{4 c^2 \ln N}{4})  = \frac{1}{N^{c^2}}
\end{align*}
\end{proof}

\begin{theorem}[Data cache capacity]
Let the total number of data requests $M \leq N^k$ for some $k > 0$.
Let $Z(\vec{0}, t)$ denote the total number of blocks in the client
data cache at time $t \in [M]$, 
assuming that the system intitally starts in the state $\vec{0}$, i.e., 
all slots are empty initially.
Then,  for $c > 0$,
\[
\Pr\left[ \max_{t\in [M]} Z(\vec{0}, t) > \sqrt{N} + 
4\sqrt{k+c} \cdot N^{\frac{1}{4}} \sqrt{\ln N}
\right] \leq 
\frac{1}{N^c}
\]
\label{thm:datacache}
\label{thm:cachecap}
\end{theorem}
\begin{proof}(\textit{sketch.})
Assume that the mixing time 
of the Markov Chain for each slot is $T$.

We show that with high probability, the data cache  
size never exceeds $\sqrt{N} + o(\sqrt{N})$
between time $T$ and $T + M$.
To show this, simply use Lemma~\ref{lem:cachesingletime}, and take 
take union bound over time steps $T$ through $T+M$:
\[
\Pr\left[ \max_{t\in [T, T+M]} Z(\vec{0}, t)  > 
\sqrt{N} + 4\sqrt{k+c} \cdot N^{\frac{1}{4}} \sqrt{\ln N} \right] 
\leq 
\frac{1}{N^c}
\]
Furthermore, notice that the cache size 
is less likely to exceed a certain upper-bound when each slot starts
empty, than starting in any other initial state. 
One can formally prove this by  
showing that for every sample path (i.e., sequence of coin flips), 
starting in the empty state
never results in more blocks in the client's data cache
than starting in any other state.

Therefore,
\begin{align*}
&\Pr\left[ \max_{t\in [M]} Z(\vec{0}, t)  > 
\sqrt{N} + 4\sqrt{k+c} \cdot N^{\frac{1}{4}} \sqrt{\ln N} \right] \\
\leq &
\Pr\left[ \max_{t\in [T, T+M]} Z(\vec{0}, t)  > 
\sqrt{N} + 4\sqrt{k+c} \cdot N^{\frac{1}{4}} \sqrt{\ln N} \right] \\
 \leq & \frac{1}{N^c}
\end{align*}
\end{proof}

\section{Recursive Construction Costs}
\label{sec:RecursiveConstructionCostsProofs}

\begin{proof}[Proof of Theorem~\ref{thm:recursive}]
The recursive O-RAM construction is obtained 
by recursively applying the O-RAM$^*$ construction $O(\log N)$
times to the  position map.  

\begin{itemize} 
\item
The amortized 
cost of the O-RAM$^*$ construction is $O(\log N)$.
The recurrence equation for the amortized cost is:
\[
T(N) = T(N/\alpha) + O(\log N) 
\]
which solves to $T(N) =  O((\log N)^2)$.
\item
The worst-case cost of the O-RAM$^*$ construction 
is $O(\sqrt{N})$. 
The recurrence relation of the worst-case cost is
\[
T(N) = T(N/\alpha) + O(\sqrt{N})
\]
which solves to $T(N) = O(\sqrt{N})$.
\item
With high probability, each partition's capacity 
in the O-RAM$^*$ construction
will not exceed $O(\sqrt{N})$ (Theorem~\ref{thm:partsizesingle}),
and hence the total server-side storage of the O-RAM$^*$ construction
is $O(N)$. 
The recurrence relation for the server-side storage is
\[
T(N) = T(N/\alpha) + O(N) 
\]
which solves to $T(N) = O(N)$. 
\item
With high probability, the client's data cache   
will not exceed $O(\sqrt{N})$ (Theorem~\ref{thm:datacache})
in the O-RAM$^*$ construction.
The recurrence relation for the client-side storage is
\[
T(N) = T(N/\alpha) + O( \sqrt{N})
\]
which solves to $T(N) = O(\sqrt{N})$.

Notice that the O-RAM construction by Goodrich \etal~\cite{MMORAM}, 
which we use for the 
partition O-RAM requires a shuffling buffer of size $O(\sqrt{N})$ for 
an O-RAM of capacity $N$. But this reshuffling buffer can be shared
across all partitions at all levels of recursion.
Therefore, it does not increase the client-side storage asymptotically.
\end{itemize} 
\end{proof}

\section{Security of the Practical Partition O-RAM}
\label{sec:practicalsecproof}
Here we prove the security of the partition O-RAM construction of our practical scheme.

\begin{lemma}
The $\pRead$ operation will never read the same block from a level more than once before that level is reshuffled.
\label{lem:ReadOnce}
\end{lemma}
\begin{proof}
A block is read from a level for one of two reasons. It is either a real block that the client wants to read from the partition, or it is a dummy block. If it is a real block, then the real block will be placed in the client's cache and the next time that same real bock is being read, it will be read from its new location. If the block is a dummy block, then after reading it, the $\cnt$ counter is incremented, ensuring that another dummy block will be read the next time a dummy must be read from that level. When a level is reshuffled, it either becomes unfilled or it is filled with a new set of blocks.
\end{proof}

\begin{theorem}
The $\pRead$ operation causes blocks to be read from the server pseudo-randomly without replacement, hence independently from the data request sequence.
\label{thm:PracticalReadPartitionSecurity}
\end{theorem}
\begin{proof}
The $\pRead$ operation reads a block from each level of the partition.  Since the blocks in a level are pseudo-randomly permuted by applying the $\prp$ function, each block read is pseudo-randomly chosen. By Lemma~\ref{lem:ReadOnce}, the block read from each level will be a previously unread block, so the blocks are read without replacement.
\end{proof}

\begin{lemma}
A reshuffling of a set of levels leaks no other information to the server except that a reshuffling of those levels occurred.
\label{lem:ReshufflingSecurity}
\end{lemma}
\begin{proof}
A reshuffling operation always reads exactly $2^\ell$ unread blocks from a level $\ell$ on the server, shuffles them on the client side, and uploads them to the first unfilled level (or the top level if all levels are filled). Since the server always knows which levels are filled, which blocks have been read, and cannot observe the shuffling that happens entirely on the client-side, then the server learns nothing new.
\end{proof}

\begin{theorem}
The $\pWrite$ operations leak no information about the data request sequence.
\label{thm:PracticalWritePartitionSecurity}
\end{theorem}
\begin{proof}
After exactly every $2^\ell$ $\pWrite$ operations, levels $0, 1, \ldots, \ell$ are reshuffled. Therefore the reshufflings happen at regular deterministic intervals. The reshufflings are the only operations caused by $\pWrite$ that are observed by the server. By Lemma~\ref{lem:ReshufflingSecurity}, the reshuffling operations leak no information to the server except that the reshufflings happened. Since the reshufflings happen at deterministic intervals independent of the data request sequence, the $\pWrite$ operations leak no information about the data request sequence.
\end{proof}

\elaine{TODO: precise sec defn for modified oram, connect all thms related to security}

\newcommand{\uniform}{\mathsf{UniformRandom}}

\section{Concurrent Constructions: Proof of Worst-Case Cost}
\label{sec:concurrentproof}
\elaine{to fix: that hoefding bound, and 
smaller buffers remain after big buffer gone}
\elaine{change the notations...}

Consider level $i$ ($1 \leq i \leq \floor{\log_{\alpha} N}$)
\elaine{explain $\alpha$}
of the recursion, let $S$ be the maximum capacity for each
partition, and let $P$ denote the number of partitions.
For the sake of asymptotic proofs, 
we focus on $1 \leq i < \frac{1}{2}\floor{\log_{\alpha} N}$ -- since
the O-RAM capacity at level 
$\frac{1}{2}\floor{\log_{\alpha} N}$ 
(or lower) is bounded by $O(\sqrt{N})$.
Therefore, we assume that the recursions stops at level
$\frac{1}{2}\floor{\log_{\alpha} N} -1$, and that 
that the client simply stores 
level $\frac{1}{2}\floor{\log_{\alpha} N}$ locally.

\elaine{do the same for other proofs to 
avoid the high probability argument for small N}

\subsection{Distribution of Amount of Work Queued for One Partition}

We now analyze the distribution of the amount of work queued 
for one specific partition at any point of time, after the system
reaches steady state.
Imagine that the amortizer schedules work fast enough, such 
that all all jobs will be dequeued in at most $\tau$ time steps.  
$\tau$ is a parameter to be determined later.

Consider the following stochastic process:
Let $C_p \in \Z_{S}$ denote a counter for partition $p$,
where $S$ is the maximum capacity of each partition.
At every time step, flip an independent random coin $r \leftarrow \uniform(1..P)$,
and let $C_r \leftarrow C_r + 1 \ (\text{mod } S)$. 

We shall first focus our analysis on a single given partition.
The stochastic process for a given partition 
can be described as below.
Basically, let $C$ denote the counter for the given partition.
At every time step, we flip an independent random coin 
which comes up heads with probability $\frac{1}{P}$.
We let $C \leftarrow C + 1 \ (\text{mod } S)$ if the coin comes up heads. 

Whenever the counter $C$ reaches a multiple of $2^i$ for some positive
integer $i$ (but $C$ is not a multiple of $2^{i+1}$),
a shuffling job of size $2^i$ is created and enqueued. \emil{Define what we mean by size. The amount of bandwidth and work performed is closer to $2 \cdot 2^i$.}
Furthermore, 
\textit{whenever a job of size $2^i$ is created and enqueued,
all existing jobs of size $\leq 2^i$ are immediately 
cancelled and
removed from the queue.}
All jobs will be dequeued after at most $\tau$ time steps. 

\begin{fact}
Consider a given partition. Suppose that jobs in the queue
are ordered according to the time they are enqueued.
Then the sizes of jobs in the queue 
are strictly decreasing.
\label{fct:jobsizedecrease}
\end{fact}

\begin{fact}
At any point of time, for a given partition, suppose 
that the largest job in the queue has size 
$2^i$, then the total size of all jobs in the queue
is bounded by $2 \cdot 2^i$.  
\label{fct:largestjob}
\end{fact}
Henceforth, we refer to the total size of all unfinished jobs
as the \textit{amount of work} for a given partition.

\elaine{explain the above more clearly...}

For a given partition, 
the above stochastic process can be modelled by a Markov Chain. 
Each state is denoted by the tuple $(c, Q)$ 
where $c$ is the counter value, and $Q$ denotes the current state
of the queue. Specifically, 
let $Q \defeq \{(s_i, t_i) \}_{i \in [\ell]}$.
$0 \leq \ell \leq \ceil{\log S}$ denotes the 
current queue length. The sequence of $s_i$'s represent
the job sizes and are strictly 
decreasing according to Fact~\ref{fct:jobsizedecrease}.
Each $0 \leq t_i < \tau$ represents the time elapsed since the $i$-th job
was enqueued.
The transitions of the Markov Chain are defined as below:

Let 
$(c, Q)$ denote the current state, where 
$Q \defeq \{ (s_i, t_i) \}_{i \in [\ell]}$. \emil{Why note define $Q'$ here and reuse it?}
\begin{itemize}
\item
If $c + 1  \ (\text{mod } S)$ is a multiple of $2^i$ 
for some positive interger $i$. 
Let $i$ denote the largest positive integer
such that $c + 1$ is a multiple of $2^i$:
\begin{itemize}
\item
With probability $\frac{1}{P}$, 
transition to state $\left(c+1 \ (\text{mod } S), Q'\right)$ where 
$Q' \defeq \{ (2^i, 0)\} \cup \{(s, t + 1) | (s, t) \in Q \text{ and }  t < \tau - 1 \text{ and } s > 2^i \}$.
\item
With probability $1 - \frac{1}{P}$, 
transition to state $\left(c, Q'\right)$ where
$Q' \defeq \{(s, t + 1) | (s, t) \in Q \text{ and }  t < \tau - 1 
\}$.
\end{itemize}
\item
If $c + 1  \ (\text{mod } S)$ is not a multiple of $2^i$:
\begin{itemize}
\item
With probability $\frac{1}{P}$, 
transition to state $\left(c+1 \ (\text{mod } S), Q'\right)$ where 
$Q' \defeq \{(s, t + 1) | (s, t) \in Q \text{ and }  t < \tau - 1 
\}$. 
\item
With probability $1 - \frac{1}{P}$, 
transition to state $\left(c, Q'\right)$ where
$Q' \defeq \{(s, t + 1) | (s, t) \in Q \text{ and }  t < \tau - 1 
\}$.
\end{itemize}
\end{itemize}
 
\elaine{may need to elaborate on this...}

It is also not hard to see that the Markov Chain is ergodic \elaine{double check},
and therefore, 
has a stationary distribution.

Let $\pi_i$ denote the probability that the largest
job in a given partition 
has size $2^i$. 
According to Fact~\ref{fct:largestjob}, the total
amount of work for that partition is then bounded by $2 \cdot 2^i$.
Since the Markov Chain has a stationary distribution, $\pi_i$ 
is both the time 
average and ensemble average in its steady state.

\begin{lemma}
The probability that the largest job in a given partition has
size $2^i$
$\pi_i = O( \frac{\tau}{P \cdot 2^i})$. \emil{Let's just put the constant in the proof statement because we use that constant later.}
\label{lem:worstcasepartdistr}
\end{lemma}
\begin{proof}
\emil{It should be mentioned here that $\tau$ will be bounded later.}
We use time average to obtain an upper bound on $\pi_i$. 
Suppose we let the Markov Chain run for a really long time $T$. 
Let $X_i$ denote the indicator random variable that the Markov Chain 
goes from $(c, *)$ to $(c+1, *)$ during time step $i \in [T]$.

Let $X(T) \defeq \sum_{i\in [T]} X_i$.
It is not hard to see that 
\[
\E[X(T)] = \frac{T}{P}
\]

Due to the central limit theorem,
as $T \rightarrow \infty$,
$\Pr[X(T) < 2\E[X(T)]] \rightarrow 1$. \emil{Where is this statement used later?}

If the counter has advanced $X$ times, then it means that the
counter has completed at most $\ceil{\frac{X}{S}}$ cycles.
In each cycle, there are at most  $\frac{S \cdot \tau}{2^i}$ \emil{where did this expression come from? I don't get it.}
time steps in which largest job  
has size $2^i$.
Therefore, over $\ceil{\frac{X}{S}}$ cycles, there are at 
most $\frac{2 X \tau}{2^i}$ 
time steps in which the largest job has size $2^i$.

Therefore, 
\begin{align*}
\forall 1 \leq i \leq \ceil{\log S}: \quad
\pi_i = \lim_{T\rightarrow \infty}\frac{1}{T}\cdot \frac{2X(T) \cdot \tau}{ 2^i }   \leq  \frac{1}{T} \cdot \frac{2 \tau} {2^i}  \cdot 2\E[X(T)]
=\frac{4\tau}{P \cdot 2^i}
\end{align*}
\end{proof}

\begin{lemma}
In steady state, the expected size
of the largest job 
for a given partition is 
upper bounded by 
$\frac{4\tau}{P} \log S$.
Moreover, the expected amount of work 
for a given partition is upper bounded by
$\frac{8\tau}{P} \log S$.
\end{lemma}
\begin{proof}
Straightforward from Lemma~\ref{lem:worstcasepartdistr}
and Fact~\ref{fct:largestjob}.
\end{proof}

\subsection{Bounding Total Amount of Work Queued for All Partitions}

\begin{lemma}[Measure concentration for independent random variables]
Let $X_1, X_2, \ldots, X_P$ denote {\it i.i.d.}
random variables. Let $\lambda \in \N$,  suppose that each $X_j$
($j \in [P]$) has non-zero probability  
for the range of integer values within $[0, 2^\lambda]$. 
Furthermore, the distribution of each $X_j$
satisfies the following.
For each integer $1 \leq i \leq \lambda$:   
$\Pr[X_j = 2^i] \leq \frac{\beta}{2^i}$
for some $\beta > 0$. \emil{Should it be noted that $\beta$ must be independant of $\lambda$?}
Let $X = \sum_{j = 1}^P X_j$. %
It is not hard to see that  
$\E[X] = P \cdot \E[X_1] \leq  \beta  \lambda P$.
Furthermore,
\[
\Pr[X \geq \beta \lambda P + 2 \cdot 2^\lambda \cdot \log \frac{\lambda}{\delta} ] \leq 
\delta
\]
\label{lem:concentrationind}
\end{lemma}

\begin{proof}
For $j \in [P]$ and $i \in [\lambda]$,  let $Y_{j, i}$ denote
the following random variable:
\[
Y_{j, i} = \begin{cases}
1 & \text{if } X_j = 2^i \\
0 & \text{otherwise}
\end{cases}
\]
Let $Y_i = 2^i \cdot \sum_{j = 1}^P Y_{j, i}$. 
It is not hard to see that $\E[Y_i] = 2^i \cdot \sum_{j = 1}^P \E[Y_{j, i}]
=  \beta P $.
Let $q = \frac{\beta}{2^i}$. 
For $\gamma \in [P]$, we have:
\begin{align*}
\Pr[Y_i \geq \gamma \cdot 2^i] & \leq {P \choose \gamma} 
\cdot q^\gamma 
\leq \frac{(q P)^\gamma}{\gamma^\gamma}  \\
\end{align*}
where the last inequality above is due to Stirling's formula. \emil{Where does the first part come from? What is the meaning of $\gamma$?}
Suppose $\gamma = q P + \epsilon$, we have:
\begin{align*}
\Pr[Y_i \geq P\beta + \epsilon \cdot 2^i] & \leq  
\left(\frac{q P}{q P + \epsilon}\right)^{q P + \epsilon}
= \left(1 - \frac{1}{(q P + \epsilon)/\epsilon}\right)^{q P + \epsilon}
\leq \exp(-\epsilon)
\end{align*}

Notice that $X = \sum_{i = 1}^\lambda Y_i$.
Therefore, due to union bound,  
\[
\Pr[X \geq \beta \lambda P +  2 \cdot 2^\lambda  \cdot \log \frac{\lambda}{\delta}] \leq
\Pr[X \geq \beta \lambda P +  \log \frac{\lambda}{\delta} \cdot \sum_{i = 1}^\lambda 2^i] \leq \delta
\]
\end{proof}

\begin{lemma}[Total amount of work over all partitions]
At any given point of time, let $X_j$ denote the 
maximum job size for the $j$-th partition.
From Lemma~\ref{lem:worstcasepartdistr},
we know that for $1\leq i \leq \ceil{\log S}$,
$1 \leq j \leq P$,
$\Pr[X_j = 2^i] \leq \frac{4\tau}{P 2^i}$,
Let $X \defeq \sum_{j \in [P]} X_j$.
Then,
\[
\Pr[X \geq 8 \tau \log S + 3 \cdot S \cdot \log \frac{\log S}{\delta} ] \leq 
\delta
\]
(Due to Fact~\ref{fct:largestjob}, the total amount of work
across all partitions is bounded by $2X$.)
\label{lem:totalwork}
\end{lemma}

\begin{proof}
Let $\beta = \frac{8\tau}{P}$, $\lambda = \log S$.
The proof is similar to that of Lemma~\ref{lem:concentrationind} \emil{do we still want to have both lemmas? What is the point of having Lemma~\ref{lem:concentrationind} when it is not actually used?},
except that the $X_j$'s are no longer independent.

We use negative associativity~\cite{negaassoc} to deal with this problem. 
For technical reasons which will become clear later,
we will consider the sum of all but $1$ partitions.  
Specifically, let $\hat X \defeq \sum_{j \in [P-1]} X_j$. 
Hence, $X = \hat X + X_P$.
Since the maximum job size for the $P\text{-th}$ partition 
can be at most $S$, it suffices to prove that
$$\Pr[\hat X \geq 8 \tau \log S + 2 \cdot S \cdot \log \frac{\log S}{\delta} ] \leq 
\delta$$

Similar to the proof of Lemma~\ref{lem:concentrationind},
for $j \in [P]$ and $i \in [\lambda]$,  let $Y_{j, i}$ denote
the following random variable, but with a minor change.
Basically, instead of letting $Y_{j, i} = 1$ when 
$X_j = 2^i$, we now let 
$Y_{j, i} = 1$ when $X_j \geq 2^i$. 
\begin{equation}
Y_{j, i} = \begin{cases}
1 & \text{if } X_j \geq 2^i \\
0 & \text{otherwise}
\end{cases}
\label{eqn:ys}
\end{equation}
Let $Y_i \defeq 2^i \cdot \sum_{j \in [P-1]} Y_{j, i}$.
Observe that in this case, $X \leq \sum_{i = 1}^\lambda Y_i$,
and $\Pr[Y_{j, i} = 1] = \Pr[X_j \geq 2^i] \leq 2 \Pr[X_j = 2^i] \leq \frac{8\tau}{P 2^i}$.

If we can prove that $Y_{1, i}, Y_{2, i}, \ldots, Y_{P-1, i}$
are negatively associated  (which we will show later), 
then we will have the following:
\begin{align*}
\Pr[Y_i \geq \gamma \cdot 2^i] & \leq 
\sum_{S \subseteq [P-1], |S| = \gamma}
\Pr[\forall j \in S: Y_{j, i} = 1 ] & \\
& = 
\sum_{S \subseteq [P-1], |S| = \gamma}
\Pr[\forall j \in S: Y_{j, i} \geq 1 ] & \\
& \leq
\sum_{S \subseteq [P-1], |S| = \gamma}
\prod_{j \in S} \Pr[Y_{j, i} \geq 1 ] & \text{(due to negative associativity, Proposition 4 in \cite{negaassoc} )}\\
& = 
\sum_{S \subseteq [P-1], |S| = \gamma}
\prod_{j \in S} \Pr[Y_{j, i} = 1 ] & \\
& = {P-1 \choose \gamma} \cdot q^\gamma 
\leq {P \choose \gamma} \cdot q^\gamma 
\end{align*}
where $q \defeq \frac{8 \tau}{P 2^i}$.
\elaine{change the notation S to something else}

From here on, the remaining proof would be similar to that of
Lemma~\ref{lem:concentrationind}.

\emil{It sounds like the part below can just be seperated into another lemma.}
So far, we have shown how to prove this lemma assuming that 
$Y_{1, i}, Y_{2, i}, \ldots, Y_{P-1, i}$ are negatively associated.
In the remainder of this proof, we will show that this is
indeed the case in the following lemma.
\end{proof}

\begin{lemma}
\label{lem:negaassocallbutone}
For any $i \in [\log S]$,
the random variables  
$Y_{1, i}, Y_{2, i}, \ldots, Y_{P-1, i}$ 
as defined in Equation~\ref{eqn:ys}
are negatively associated.
\end{lemma}
\begin{proof}
Suppose we run for $M$ time steps -- $M$ is
large enough such that the system reaches steady state, 
and we examine the
random variables $Y_{j,i}$'s
at the end of $M$ time steps. 

To show that after $M$ time steps, 
$Y_{1, i}, Y_{2, i}, \ldots, Y_{P-1, i}$ 
are negatively associated, we first argue that 
after $M-\tau$ steps, the counters of each partition $C_1, C_2, \ldots,
C_{p-1}$ (mod $S$) are independent.
To see this, think of the vector 
$(c_1, c_2, \ldots, c_{P-1})$
as the configuration of counters of
all but one partitions at time $t$.
From time $t$ to $t+1$, the configuration vector
$(c_1, c_{2}, \ldots, c_{i-1}, c_{i} + 1 (\text{mod }S), c_{i+1}, \ldots,
c_{p-1})$ 
can transition into any neighboring vector
with probability $\frac{1}{P}$, and 
self-loops with probability $\frac{1}{P}$.
It is not hard to see that configuration vector forms
an ergodic Markov Chain.
In addition, in the stationary distribution,  
all configurations are ``equivalent'', 
so all configurations have the same probability.
It is not hard to see now that the counters
$(C_1, \ldots, C_{P-1})$ are independent random variables
at time step $M-\tau$.
Note that the reason why we left one partition out is
because the counters  
the counters $(C_1, \ldots, C_P)$ form a \textit{periodic} Markov 
Chain (i.e., the sum of all counters should be equal
to the number of time steps mod $S$), 
therefore, a stationary distribution does not exist. 

We now argue why  after $M$ time steps,
$Y_{1, i}, Y_{2, i}, \ldots, Y_{P-1, i}$ 
are negatively associated. 

For $j \in [P-1], t \in [M-\tau + 1, M]$, 
let $B_{j, t}$ be indicator random variable as defined below:
\[
 B_{j, t} \defeq \begin{cases}
1 &  \text{partition $j$ is chosen in the $t$-th time step \emil{Chosen for what? We should be specific.}} \\
0 &  \text{otherwise} \\
\end{cases}
\]
Due to Proposition 11 of \cite{negaassoc}, the vector 
$\{B_{j, t}\}_{j \in [P-1], t \in [M-\tau +1, M]}$ 
is negatively associated.
Earlier, we argued that the partition counters mod $S$ 
for all but one partition are
indepdent. Using a similar argument,
we can prove that the partition counters 
$(C_1, C_2, \ldots, C_{P-1})$ (mod $2^i$)
are independent.
Since the variables 
$\{B_{j, t}\}_{j \in [P-1], t \in [M-\tau +1, M]}$ 
are independent of the partition counters 
$(C_1, C_2, \ldots, C_{P-1})$ (mod $2^i$)
at time $M-\tau$,
we can conclude that 
the union of the partition counters 
$(C_1, C_2, \ldots, C_{P-1})$ (mod $2^i$)
and the vector 
$\{B_{j, t}\}_{j \in [P-1], t \in [M-\tau +1, M]}$ 
is negatively associated.
Without loss of generality, 
we assume that every job will remain in the queue for
exactly $\tau$ amount of time, even if the job may be completed
before that. 

At the end of $M$ time steps, the random
variable $Y_{j, i}$ is a \textit{non-decreasing} 
function of the partition counter $C_j$ (mod $2^i$) at time $M -\tau$, 
and $\{B_{j, t}\}_{t \in [M-\tau+1, M]}$. 
Due to Proposition 7 of \cite{negaassoc},
the vector $(Y_{j, i})_{j \in [P-1]}$ is negatively associated.
Furthermore, this holds for all $i \in [\lambda]$. \emil{Is the concatenation of all of those vectors also negatively associated? We should state a specific lemma from the other paper.}
\end{proof}

\paragraph{Amount of work amortized to each time step}
We will pick $\tau = P = \sqrt{N}$.
As a result, if the amortizer schedules $16 \log S$ 
work to be done per time step, then with 
$1 - \frac{1}{\text{poly}(N)}$ probability,
we guarantee that each job remains in the queue for 
$\tau$ time or less. 

The remainder of this section will bound the 
additional client cache required
for the concurrent construction, including the memory
necessary for storing cached reads and pending writes.

\subsection{Bounding the Size of Cached Reads}
For the practical construction, if a partition currently 
has a shuffling job up to level $\ell$ in the queue, 
and a $\pRead$ operation occurs,
then the blocks read from levels $1, \ldots, \ell$ of that partition
needs to be cached, since the same block cannot be read twice.
Note that reads for levels $\ell +1$ and higher need not be cached.  
We now bound the size of client cache needed to store these
cached reads. 

Note also that in the theoretic construction 
the buffer for cached reads is optional -- since we use a variant of the 
Goodrich-Mitzenmacher~\cite{MMORAM} O-RAM as the partition O-RAM,
where we do not use any dummy blocks, and the O-RAM can be read
an unlimited number of times. In particular, every time we read
a level, it always appears to the server that two random locations
are being accessed.

Without loss of generality, 
we will assume that every cached read will remain in the client's cache
for $\tau$ amount of time. In reality, since all shuffling jobs are  
guaranteed to be done within $\tau$ time, each cached read
will remain in the cache for $\tau$ time or less.
Therefore, this allows us to obtain an upper-bound on the
size of cached reads in the client's cache.

We first study the distribution of the size of cached reads for 
a single partition.
At some time $t + \tau$, the cached reads are accumulated
from time steps $t$ to $t + \tau$. 
For each cached read, its size is the 
the logarithm of the size of the largest job at the time 
the cached read is created. \emil{Explain this better. Say why.}
For the sake of an upper bound,
we will assume that within a window $[t, t + \tau]$, the size
of all cached reads are the logarithm of the size of the largest job 
between time $t$ and $t + \tau$.
 
\begin{lemma} 
In steady state,
let $p_i$ denote the probability that the largest job between
a time window $[t, t+\tau]$ for a given partition
is of size $2^i$, for any time $t$.
Let $\tau = P$.
Then,
\[
p_i \leq \frac{8}{2^i}
\]
\end{lemma}
\begin{proof}
\emil{Why $8$?}
Consider the Markov Chain for the counter $c$ for a given partition.
\elaine{expand the MC description}

Let $q_c$ denote the probability that the counter is $c$ mod $S$
in the steady state.
It is not hard to see that $q_c = 1/S$ for any $1 \leq c \leq S$.
\elaine{the mod S notation is a bit problematic at S. throughout}.

Let $R_t$ be a random variable representing the counter value
at time $t$.

\begin{align*}
 &\Pr[\text{largest job between $t$ and $t+\tau$ has size $2^i$}] \\
\leq & 
\sum_{0\leq \alpha\leq 2\tau} 
\Pr[\exists k: R_{t-\tau} = k \cdot 2^i - \alpha] 
\cdot \Pr[\text{this partition
is selected $\geq \alpha$ times in a time window of $2\tau$}]\\
\leq &\sum_{0 \leq \alpha \leq 2\tau} \frac{1}{2^i} \cdot
{2\tau \choose \alpha} \cdot \frac{1}{P^\alpha}\\
\leq & \sum_{0 \leq \alpha \leq 2\tau} \frac{1}{2^i} \cdot \exp(-(\alpha - \frac{2\tau}{P}))
\end{align*}

When we select $\tau = P$, it is not hard to see that

\begin{align*}
\Pr[\text{largest job between $t$ and $t+\tau$ has size $2^i$}]  
\leq \frac{8}{2^i} 
\end{align*}
\elaine{make sure that constant 8 works}
\end{proof}

\begin{lemma}
At any point of time $t$ in steady state,
let $X_j$ denote the logarithm 
of largest job size for partition $j \in [P]$ in
a time window $[t, t+\tau]$.
Let $X = \sum_{j \in [P]} X_j$.
Then
\[\Pr[ X \geq \E[X] + (\sqrt{P} + 1) \log S \sqrt{\log \frac{1}{\delta}}] \leq 
\delta \]
\label{lem:cachedreadssumparts}
\end{lemma}
\begin{proof}
For technical reasons, we first bound
the sum of $X_j$'s for all but one partitions, i.e., 
$X_1, \ldots, X_{P-1}$.
Let $\hat X = \sum_{j = 1}^{P-1} X_j$.
Since $X_P \leq \log S$, 
it suffices to prove that
\[\Pr[ \hat X \geq \E[X] + \sqrt{P}  \log S \sqrt{\log \frac{1}{\delta}}] \leq 
\delta \]
We first show that $X_1, X_2, \ldots, X_{P-1}$'s 
are negatively associated,
and we then apply Hoeffding's inequality which holds
for negatively associated random variables. 
The proof of negatively associativity is similar to the proof of 
negative associativity in Lemma~\ref{lem:negaassocallbutone}, assuming
that all cached reads will remain in the cache for 
exactly $\tau$ time. 

Now, according to Hoeffding inequality,
\begin{align*}
\Pr[ \hat X \geq \E[X] + \epsilon P] \leq 
\exp\left(-\frac{2 \epsilon^2 P^2}{P (\log S)^2}\right) 
= 
\exp\left(-\frac{2 \epsilon^2 P}{(\log S)^2}\right) 
\end{align*}
Let $\epsilon =\frac{\log S}{\sqrt{P}} \sqrt{\log \frac{1}{\delta}}$,
we have 
\[\Pr[ \hat X \geq \E[X] + \sqrt{P} \log S \sqrt{\log \frac{1}{\delta}}] \leq 
\delta \]

When $\tau = P$, 
$\Pr[X_j = i ] \leq \frac{8}{2^i}$ for all $j \in [P]$.
Therefore,
it is not hard to see that $\E[X] \leq 16 P$.
Therefore,
\[\Pr[ \hat X \geq 16P + \sqrt{P} \log S \sqrt{\log \frac{1}{\delta}}] \leq 
\delta \]
\end{proof}

We now bound the size of cached reads at any point of time 
in steady state.
The cached reads at time $t+\tau$ are created
during time steps $t$ through $t+\tau$.
Therefore, it suffices to bound the total size of cached
reads created during time steps $t$ through $t+\tau$.

Let $\vec{X} \defeq (X_1, X_2, \ldots, X_P)$ denote 
the logarithm of the maximum
job size for each partition, 
during time steps $t$ through $t+\tau$.
If time step $t + i$ where $0 \leq i \leq \tau$  
chooses partition $j$, then $X_j$ number of cached 
reads are created.

\begin{lemma}
Supose we are given that $\vec{X} = \vec{x}$, and let 
$\mu = \text{mean}(\vec{x})$.
For $0\leq i \leq \tau$, 
let $Y_i$ denote the number of cached reads created 
in time step $t+i$.
Let $Y \defeq \sum_{i = 0}^\tau Y_i$.
\[
\Pr[Y \geq \tau \mu + \sqrt{\tau} \cdot \log S \cdot \sqrt{\log \frac{1}{\delta}} \ | \vec{X} = \vec{x}] \leq \delta
\]
\label{lem:cachedreadsconditionprob}
\end{lemma}
\begin{proof}
For at time $t+i$ (where $i \in [0, \tau]$)
suppose partition $j$ is chosen.
Recall that we assume the number of cached reads
the number of cached reads $Y_i$ generated in that time step
depends on the largest job size
of partition $j$ between the entire $[t, t+ \tau]$.
As a result, given a fixed $\vec{X}$, 
each $Y_i$
(where $i \in [0, \tau]$) depends only on which partition is
chosen to be read in time step $t + i$.
Therefore, the $Y_i$'s are independent.

By Hoeffding bound,
\begin{align*}
\Pr[Y \geq \tau \mu + \tau \epsilon | \vec{X} = \vec{x}] &\leq
\exp\left(-\frac{2 \epsilon^2 \tau^2}{\tau (\log S)^2}\right)\\
&=\exp\left(-\frac{2 \epsilon^2 \tau}{(\log S)^2}\right)\\
\end{align*}
Let $\epsilon = \frac{\log S}{\sqrt{\tau}}\sqrt{\log \frac{1}{\delta}}$,
we have the following:
\begin{align*}
\Pr[Y \geq \tau \mu + \sqrt{\tau} \cdot \log S \cdot \sqrt{\log \frac{1}{\delta}} \ | \vec{X} = \vec{x}] \leq \delta
\end{align*}
\end{proof}

\begin{theorem}[Bounding cached reads]
Consider the system in steady state.
For $0\leq i \leq \tau$, 
let $Y_i$ denote the number of cached reads created 
in time step $t+i$.
Let $Y \defeq \sum_{i = 0}^\tau Y_i$.
Assume that $\tau = P$.
Then,  %
\[
\Pr\left[Y \geq  16P + 3 \sqrt{P} \log S \sqrt{\log \frac{1}{\delta}} 
\right] \leq  2\delta 
\]
\end{theorem}
\begin{proof}
According to Lemma~\ref{lem:cachedreadssumparts}, 
\[\Pr[ \text{mean}(\vec{X}) 
\geq  16  + \frac{2\log S}{\sqrt{P}} \sqrt{\log \frac{1}{\delta}}] \leq 
\delta \]
Let $G$ denote the event that 
$\text{mean}(\vec{X}) 
\leq 16  + \frac{2\log S}{\sqrt{P}} \sqrt{\log \frac{1}{\delta}}$.
Let $\bar{G}$ denote the event that
$\text{mean}(\vec{X}) 
>  16 + \frac{2\log S}{\sqrt{P}} \sqrt{\log \frac{1}{\delta}}$.
Due to Lemma~\ref{lem:cachedreadsconditionprob}, 
\[
\Pr\left[Y \geq  16\tau + \frac{ 2 \tau \log S}{\sqrt{P}} \sqrt{\log \frac{1}{\delta}} +  \sqrt{\tau} \cdot \log S \cdot \sqrt{\log \frac{1}{\delta}} \ | G 
\right] \leq \delta 
\]

Therefore,
\begin{align*}
&\Pr\left[Y \geq 16\tau+ \frac{ 2\tau \log S}{\sqrt{P}} \sqrt{\log \frac{1}{\delta}} +  \sqrt{\tau} \cdot \log S \cdot \sqrt{\log \frac{1}{\delta}} 
\right] \\ \leq &
\Pr\left[Y \geq  16\tau + \frac{ 2\tau \log S}{\sqrt{P}} \sqrt{\log \frac{1}{\delta}} +  \sqrt{\tau} \cdot \log S \cdot \sqrt{\log \frac{1}{\delta}} \ | G
\right] + \Pr[ \bar{G}]
\leq 2\delta 
\end{align*}
\end{proof}
Plug in $\tau = P$, we get the above theorem.

\subsection{Bounding the Size of Pending Writes}
In the concurrent scheme, when a $\pWrite$ operation occurs,
the block to be written is added to the pending write queue. 
At each time step, $O(1)$ number of pending writes are created.
Since we guarantee that all queued shuffling jobs 
are completed in $\tau$ time, each pending write will be completed
in $\tau$ time.
Therefore, the total number of pending writes in the system is bounded
by $O(\tau)$.

\end{document}